\documentclass[lettersize,journal]{IEEEtran}
\usepackage{amsmath,amsfonts}
\usepackage{algorithmic}
\usepackage{algorithm}
\usepackage{array}
\usepackage[caption=false,font=normalsize,labelfont=sf,textfont=sf]{subfig}
\usepackage{textcomp}
\usepackage{stfloats}
\usepackage{url}
\usepackage{verbatim}
\usepackage{graphicx}
\usepackage{newtxtext}
\usepackage[T1]{fontenc}
\usepackage{cite}
\usepackage{multirow} 
\usepackage{siunitx}

\usepackage{placeins}

\usepackage{balance}

\usepackage{array}
\newcolumntype{P}[1]{>{\centering\arraybackslash}p{#1}}
\usepackage{tabularx}
\usepackage[british]{babel}

\usepackage{paralist}  
  \setdefaultleftmargin{1em}{1.5em}{1em}{}{}{}

 \usepackage[compact]{titlesec} 
    \titlespacing{\subsection}{0ex}{1.5ex}{0.5ex}

\usepackage{amsmath}

\usepackage{graphicx}
\usepackage{colortbl}
\usepackage{hhline}
\usepackage{pifont}
\usepackage[justification=centering]{caption}

\usepackage[normalem]{ulem}
\useunder{\uline}{\ul}{}

\usepackage{array}

\usepackage{comment}

\usepackage{graphicx}
\usepackage{color}
\definecolor{greytext}{gray}{0.5}
\definecolor{DarkGreen}{rgb}{0.0, 0.5, 0.0}
\definecolor{DarkKhaki}{rgb}{0.74, 0.72, 0.42}
\definecolor{DarkRed}{rgb}{0.7, 0.2, 0.2}
\definecolor{Purple}{rgb}{0.7,0.0,0.7}
\definecolor{Orange}{rgb}{0.9,0.45,0.31} 
\definecolor{Teal}{rgb}{0.12,0.5,0.5}
\definecolor{Black}{rgb}{0,0,0}
\definecolor{DeepPink}{rgb}{1,0.08,0.34}
\definecolor{Mustard}{rgb}{0.8, 0.6, 0}
\definecolor{Banana}{rgb}{0.996, 0.867, 0}

\usepackage{gensymb}

\definecolor{fuchsiapink}{rgb}{1.0, 0.47, 1.0}

\newcommand{\rev}[1]{#1}

\definecolor{kmCGreen}{rgb}{0.4, 0.6, 0.6}
\definecolor{kmEGreen}{rgb}{0.0, 0.5, 0.24} 
\definecolor{kmGreenbright}{rgb}{0.10, 0.8, 0.443}  

\newcommand{\eg}{\textit{e.g.,~}}
\newcommand{\ie}{{\textit{i.e.},~}}
\newcommand{\vs}{\textit{vs.~}}

\newcommand{\etal}{\textit{et al}}

\newcommand{\lowHeading}[1]{\vspace{0.5pt}\textit{#1:}}
\newcommand{\lowHeadingNoPunct}[1]{\vspace{0.5pt}\textit{#1}}

\newcommand{\tabrowskip}[0]{\vspace{2.5pt}}
\def\arraystretch{1.0}

\usepackage{csquotes}  

\usepackage{dirtytalk} 

\usepackage{booktabs}
\usepackage{wrapfig}
\usepackage{tabularx}
\usepackage[caption = false]{subfig}

\newcommand{\chora}{\textsc{chora}}

\newcommand{\ABS}{\textsc{abs}}
\newcommand{\INA}{\textsc{ina}}
\newcommand{\OPT}{\textsc{opt}}
\newcommand{\REF}{\textsc{ref}}
\newcommand{\dINA}{$\Delta_{\INA}$}
\newcommand{\dOPT}{$\Delta_{\OPT}$}
\newcommand{\dREF}{$\Delta_{\REF}$}

\newcommand{\GSR}{\textsc{gsr}}
\newcommand{\HR}{\textsc{hr}}
\newcommand{\RR}{\textsc{rr}}

\newcommand{\qSTAI}{\textsc{stai-6$_{Q}$}}
\newcommand{\qSAM}{\textsc{sam$_{Q}$}} 
\newcommand{\qSAMV}{\textsc{sam-v$_{Q}$}}
\newcommand{\qSAMA}{\textsc{sam-a$_{Q}$}}
\newcommand{\qSAMD}{\textsc{sam-d$_{Q}$}}

\newcommand{\qER}{\textsc{er$_{Q}$}} 

\newcommand{\qBFI}{\textsc{bfi-10$_{Q}$}} 
\newcommand{\qDASS}{\textsc{dass-21$_{Q}$}}  
\newcommand{\qGODSPEED}{\textsc{godspeed$_{Q}$}}

\newcommand{\qDEMO}{\textsc{demo$_{Q}$}} 
\newcommand{\qCERF}{\textsc{er}-\chora$_{Q}$} 
\newcommand{\qAOO}{\textsc{aoo$_{Q}$}}
\newcommand{\qIDR}{\textsc{idr$_{Q}$}}

\begin{document}

\title{Haptically Experienced Animacy Facilitates Emotion Regulation: A Theory-Driven Investigation}

\author{Preeti Vyas$^1$, Bereket Guta$^1$, Tim G. Zhou$^1$, Noor Naila Himam$^1$, Andero Uusberg$^2$, and Karon E. MacLean$^1$
\thanks{*This work was supported in part by Natural Sciences and Engineering Council of Canada (NSERC).
$^{1}$Preeti Vyas, Bereket Guta, Tim G. Zhou, Noor Naila Himam, and Karon E. MacLean are with the Faculty of Science, Department of Computer Science,
        University of British Columbia, Vancouver, Canada.
        {\tt\small pv@cs.ubc.ca}
$^{2}$Andero Uusberg is with the Institute of Psychology, University of Tartu, Tartu, Estonia}%
}

\markboth{Journal of \LaTeX\ Class Files,~Vol.~XX, No.~XX, June~2025}%
{Shell \MakeLowercase{\textit{et al.}}: A Sample Article Using IEEEtran.cls for IEEE Journals}

\maketitle

\begin{abstract}
Emotion regulation (ER) is essential to mental well-being but often difficult to access, especially in high-intensity moments or for individuals with clinical vulnerabilities. 
While existing technology-based ER tools offer value, they typically rely on self-reflection (e.g., emotion tracking, journaling) or co-regulation through verbal modalities (reminders, text-based conversational tools), which may not be accessible or effective when most needed. 
{The biological role of the touch modality makes it an intriguing alternate pathway 
, but}
empirical evidence is limited and under-theorized.
Building on our prior theoretical framework describing how a comforting haptic co-regulating adjunct (\chora) can support ER, we developed a zoomorphic robot \chora\ with looped biomimetic breathing and heartbeat behaviors. 
We evaluated its effects in a mixed-methods in-lab study (N=30), providing physiological, self-report, custom questionnaire, and retrospective interview data. 
Our findings demonstrate the regulatory effects of haptically experienced animacy, corroborate prior work, and validate \chora’s {theoretically grounded} potential to facilitate four ER strategies.
\end{abstract}


\section{Introduction}
\label{1:intro}

Mental health challenges are a growing global concern, with emotional dysregulation central to conditions such as anxiety, depression, and stress-related disorders~\cite{world2022world}.
\textit{Emotion regulation (ER)}, which comprises many ways of monitoring, evaluating, and modifying our emotional reactions, form the basis for human coping, resilience, and general well-being~\cite{gross1998:emoRegReview}. 
In-the-moment application of adaptive (helpful in the long term) ER strategies can be difficult, especially during intense emotional experiences or for individuals with clinical vulnerabilities~\cite{sachs-ericsson_cognitive_2021}, so
accessing their full potential benefits from external support~\cite{gross1998:emoRegReview}.
While technology tools show promise, current solutions are mostly text-based and suited for offline use~\cite{slovak2023designing}, which limits their effectiveness in high-stress moments.

Prior works demonstrate the potential of touch interaction to influence physiology and emotional state, in both human-human and human-robot contexts~\cite{eckstein_calming_2020, vyas_descriptive_2023, vyas:2024:happraisalModel}.
Humans often attribute comfort, safety, and emotional connection to touch interactions, from hugging a loved one or a pet to holding a cherished object~\cite{cascio2019social}. 
Neuroscience and psychology findings elucidate the interconnections among touch, emotion, and cognition~\cite{eckstein_calming_2020}.
Touch also supports intuitive, embodied, and non-verbal interaction, minimizing the cognitive load of text-based or conversational tools~\cite{slovak2023designing}.
However, existing evidence does not illuminate physiological and cognitive pathways and mechanisms 
\rev{of touch iteractions facilitating ER} 
due to limited focus and theorization~\cite{vyas_descriptive_2023, zhou_tangible_2024}. 

We previously proposed a theoretical framework~\cite{vyas:2024:happraisalModel} through which a ``\textbf{c}omforting, \textbf{h}aptic c\textbf{o}-\textbf{r}egulating \textbf{a}djunct'' (\chora) can help individuals deploy four established ER strategies: 
    \textit{situation selection and modification},
    \textit{attention deployment}, 
    \textit{cognitive change}, and 
    \textit{response modulation}~\cite{gross1998:emoRegReview}.
{We reasoned that a \chora\ 
with animate behaviors
would have an enhanced ability to soothe, capture attention, act as an ally, and eventually facilitate cognitive change.
\rev{As a primary objective, here we sought to empirically validate this theoretical framework using a simple robot with a zoomorphic form and motion as a \chora, previously shown to be emotionally evocative and convey animacy\cite{cang_cuddlebits_2015}.}


\rev{Sefidgar \etal~\cite{sefidgar_design_2016}} found that interaction with a `breathing' (\vs stationary) robot reduced physiological correlates of stress, an example of facilitated \textit{response modulation}.
\rev{As this study aligns most closely with our research premise, we also investigated, as a {first step and} secondary objective, whether the demonstrated physiological calming effect extends to our current \chora\ prototype (a substantially modified design w.r.t.~\cite{sefidgar_design_2016}, see Appendix G).
Building on this, we then examined \chora’s facilitation of all four ER strategies.}


    \begin{figure}[t]
    \centering
    \includegraphics[width=0.8\columnwidth]{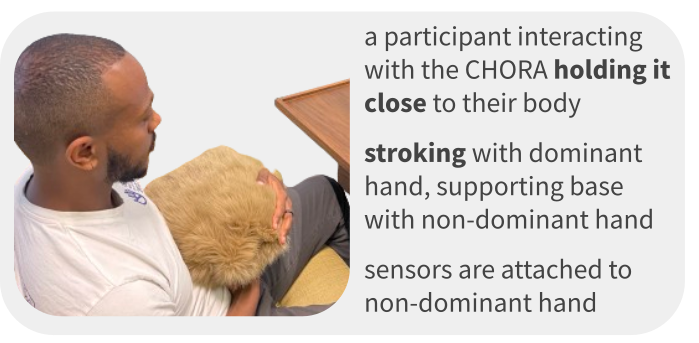}
    \caption{A participant interacting with the \chora.}
    \label{fig:s1:chora}
     \vspace{-18pt}
    \end{figure}

Our \chora, a zoomorphic robot based on the CuddleBits platform~\cite{cang_cuddlebits_2015},
used two parameterizations of two animating behaviors (breathing and heartbeat):
   one drawn from our proposed design principles~\cite{vyas:2024:happraisalModel} and refined for optimal calming effects (parameter setting pilot, N=10), 
    and the other from~\cite{sefidgar_design_2016}.
Our primary in-lab study (N=30) used physiological, self-report, and interview data to 
    evaluate the \chora's ER facilitation 
    and 
    impact of the two behavior parameterizations compared with an inanimate 
    \chora.
We contribute:
    
\begin{compactdesc}
    \item[\rev{{Primary:}} Evidence of \chora-Linked ER Facilitation] \rev{{grounded in empirical validation of}
    our theoretical framework~\cite{vyas:2024:happraisalModel}, providing converging evidence from physiological measurements, standard and custom self-report scales, and qualitative interview analyses.}
    
    \item[\rev{{Secondary:}} Corroboration of past findings~\cite{sefidgar_design_2016}] \rev{on the effects of robot behavior on physiology and emotional state, 
    observed with a substantially modified robot design and behavior parameters, {documenting the robustness of these effects at least across these design variations.}}
  
\end{compactdesc}

    %


    

\section{Related Work}
\label{2:rw}
This work is situated in the growing field of {affective haptic system design}~\cite{vyas_descriptive_2023} 
and adds to its theoretical foundations
at the intersection of psychology and interaction design. 

\subsection{Strategies 
for Emotion Regulation}
\label{2:rw:grossER}

Gross’s {well-established ER} process model}~\cite{gross1998:emoRegReview}) identifies four families of ER strategies that individuals employ to regulate their emotions: 
    \textit{situation selection and modification}   
        {(shaping one’s environment or interactions to influence emotions)},
    \textit{attentional deployment} 
        {({re}directing attention to influence one’s emotional experience, \eg distraction)}; 
    \textit{cognitive change}  
        {(modifying one's appraisal or interpretation to alter the situation's emotional significance \eg reappraisal)}; and 
    \textit{response modulation}  
        {(influencing experiential, behavioral, or physiological aspects of an emotional response once it has been generated, \eg suppression, relaxation)}. 
Individuals might use these deliberately; they can also unfold involuntarily through automatic processes. 
A strategy's adaptiveness depends on how well it supports one's long-term goals in a given emotional context.
{Adaptive} 
strategies like reappraisal tend to be both effective in the short term and healthy in the long term~\cite{uusberg_reappraising_2019}.
In contrast, maladaptive strategies such as suppression or rumination may offer short-term relief but tend to amplify distress in the long run.

\textit{Co-regulation} (where one person or agent helps regulate another’s emotional state through supportive interaction) can scaffold the often-difficult application of adaptive regulation~\cite{uusberg_reappraising_2019}.
{Technological co-regulation agents might support adaptive ER
{particularly where human support is limited or unavailable ~\cite{eckstein_calming_2020, zhou2025squeeze, sumioka_huggable_2013}.

\subsection{Why Haptic Interaction for Emotion Regulation?}
\label{2:rw:whyhaptic}
Emotion regulation has been technologically facilitated through tracking, reminders to perform pre-defined actions, prompt-based journaling, and emotionally supportive messages, \eg from chatbots~\cite{slovak2023designing}.
These approaches often require verbal engagement with a device, which can be impractical during intense or social moments.
Touch could be a more accessible, immediate, and emotionally resonant modality~\cite{eckstein_calming_2020}.

Touch interaction with a human, animal, or even a robot can shape emotions in diverse ways and across different time scales~\cite{eckstein_calming_2020}.
For ER, we focus on affiliative (positive, nurturing) touch manifested in caregiving and social bonding~\cite{kryklywy_characterizing_2023}.
While most studied in human or animal interactions, similar emotional benefits could arise with non-living tangible objects, given trust or a mental projection of animacy~\cite{eckstein_calming_2020}.

{Physiologically}, affiliative touch stimulates specific skin receptors (namely C-tactile afferent fibres), triggering oxytocin release~\cite{kryklywy_characterizing_2023},
promoting sensory calming, positive affect, and prosocial behavior~\cite{eckstein_calming_2020} \rev{and reducing perception of pain ~\cite{yim_wearable_2022, geva_touching_2020}}.
{Cognitively}, incidental haptic sensations can influence higher-order judgments, such as perceptions of people or situations
    ~\cite{ackerman_incidental_2010}.
More centrally, affiliative touch is proposed to act as a safety signal by inhibiting the amygdala’s alertness response, restoring access to cognitive processes otherwise dampened in fight-or-flight states and providing further sensorial calming {{\cite{eckstein_calming_2020} 
}.


Thus, touch is involved in ER both through direct physiological mechanisms and cognitive and relational pathways. 
This informs our exploration of exactly how haptically enriched interactions can evoke affiliative, calming, and co-regulatory responses~\cite{vyas:2024:happraisalModel} to facilitate {ER.}

In the current landscape of affective haptics system design}~\rev{\cite{vyas_descriptive_2023, zhou_tangible_2024}}, many employ vibrotactile stimuli to provide embodied ER cues, \eg by rendering a “calm” heartbeat for the user to mirror,
or to guide practices such as meditation
and controlled breathing~\cite{vyas_descriptive_2023}. 
However, these do not harness the potentially deeper connection of touch stimuli, which engage both the brain’s cognitive and social networks and its sensory and autonomic systems ~\cite{eckstein_calming_2020}.
Existing haptic technologies rarely draw explicitly on the diverse neuropsychology of touch to go beyond surface-level guidance cues.
  
\subsection{\chora: A Broad Class of Haptic ER Support Technology}
\label{2:rw:chora}

In prior work~\cite{vyas:2024:happraisalModel}, we identified a class of \chora\ (comforting, haptic co-regulating adjunct) technology that encompasses a range of forms and behaviors.
Whereas most existing examples facilitate \textit{response modulation} 
    (targeting users’ physiological states through direct, tangible cues~\cite{vyas_descriptive_2023, zhou_tangible_2024},  
    \eg Paro~\cite{geva_touching_2020}, 
    Haptic Creature~\cite{sefidgar_design_2016}, HuggieBot~\cite{block_softness_2019},
{we argue that touch interaction may have a broader regulatory potential, \rev{\ie by facilitating all four ER strategies}.} 
{The  theoretical framework proposed in \cite{vyas:2024:happraisalModel} aligns \chora\ technology} with Gross’s ER process model (\S\ref{2:rw:grossER}), articulating its capacity to \textit{haptically} support 
the four families of ER strategies:     
    
\begin{compactdesc} 
    \item[{Response {modulation:}}]  physiological calming {through haptically perceived form, texture, or behaviors};
    \item [{Attention {deployment:}}] redirection from threats or rumination {when behaviors capture and hold attention};
    \item[{Situation {modification:}}] {altered by recruiting a pro-social, animate, co-regulating} ally;
    \item[{Cognitive {change:}}] triggering associations which activate alternative appraisals. 
\end{compactdesc}
    
\rev{In this work, we sought to empirically validate whether and how an animate \chora\ facilitates all four ER strategies that contribute to emotion down-regulation, in contrast to past work that primarily validated \textit{response modulation}. 
This theory-grounded investigation, scaffolded by {the} \chora\ ER Framework~\cite{vyas:2024:happraisalModel}, advances a mechanistic 
understanding of how affective haptic technologies exert their regulatory effects, moving beyond 
{empirical} demonstrations that they elicit a general positive emotional impact {such as}~\cite{sefidgar_design_2016}.}

\subsection{\rev{The Case for an Animate \chora}}
\label{2:rw:animate}

A \chora\ can vary in appearance, form factor, interactive behaviors, and sensing capability. 
An animate form and behaviors~\rev{\cite{adachi_development_2024, geva_touching_2020, sefidgar_design_2016, sumioka_huggable_2013, yim_wearable_2022}} {are particularly relevant to ER~\cite{vyas:2024:happraisalModel}.}
\rev{Animate behaviors can be incorporated in a \chora\ in varied ways: \eg by simulating physiological rhythms, such as vibrotactile heartbeats~\cite{sumioka_huggable_2013, borgstedt_soothing_2024}, breathing~\cite{sefidgar_design_2016} or pneumatic inflation to mimic hand-holding~\cite{yim_wearable_2022}; implementing active social gestures, such as head bunting~\cite{adachi_development_2024} or responsive wiggling~\cite{geva_touching_2020} and providing reactive tactile affordances, such as huggability~\cite{ sumioka_huggable_2013} and squeezability~\cite{zhou2025squeeze}.}

\rev{In this work, we prioritized \textit{breathing} and \textit{heartbeat} as 
primary 
actuation modalities to evoke animacy. 
Unlike active social gestures (\eg head bunting) which invite stimulation, or reactive affordances (\eg squeezability) that require physical engagement from the user, these basal physiological rhythms offer a mechanism 
for \textit{passive co-regulation}. 
We posit that these cues are uniquely suited to facilitate \textit{affective entrainment}, providing a consistent rhythmic anchor that guides the user’s own physiology toward a state of rest without imposing a cognitive or physical load.}

\subsection{Breathing \& Heartbeat \rev{to Evoke Animacy}}
\label{2:rw:behavior}
\lowHeading{Breathing}
Controlled slow, deep breathing activates the parasympathetic nervous system, reducing physiological arousal and negative emotions such as anxiety and stress~\cite{jerath2015self}.
Focusing attention on the breath, a central practice in {mindfulness meditation}, also decreases activation in the amygdala (a brain region central to emotion processing) and enhances prefrontal regulation during aversive  experiences~\cite{kral2018impact}.
Different emotions are associated with distinct breathing patterns; \eg rapid, shallow breathing often accompanies anxiety, while slow, deep breathing is linked to calmness~\cite{ashhad2022breathing}. 
Passive, {haptic perception of}  a robot's slow, rhythmic ``breathing'' has been shown to facilitate nonvolitional physiological calming ~\cite{sefidgar_design_2016}.
{More recently, researchers have created haptic tools to help users {intentionally} practice and achieve slow rhythmic breathing to influence their emotional state~\cite{vyas_descriptive_2023}.

\begin{figure*}[htb!]
    \centering
    \includegraphics[width=\textwidth]{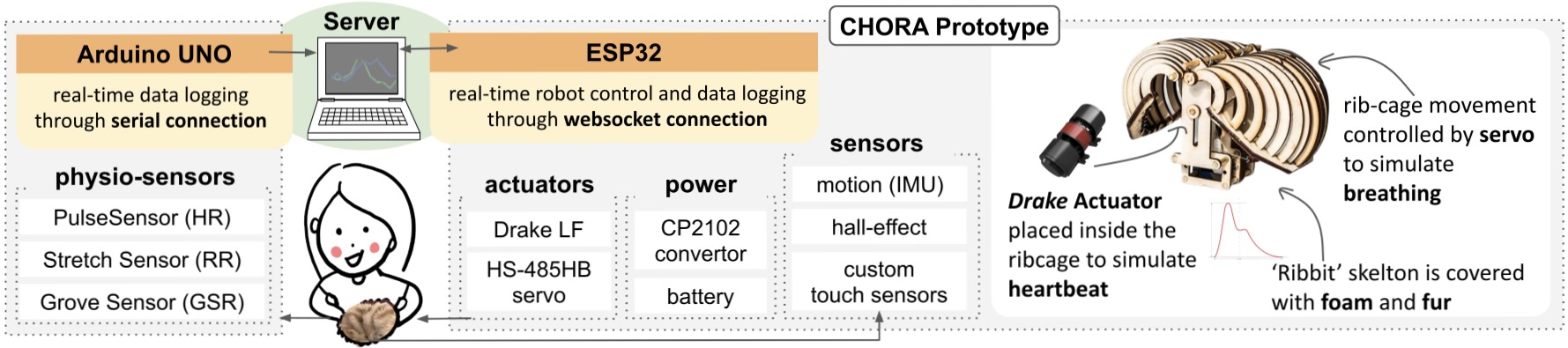}
    \caption{Research test-bed architecture with components of the \chora.} 
    \label{fig:s3:researchTestbed}
     \vspace{-15pt}
\end{figure*}

\lowHeading{Heartbeat}
Passively perceiving a calm, regular {beat} delivered through haptic cues can override users’ perception of their own heart rate, promoting emotional entrainment and reducing anxiety without disrupting ongoing tasks~\cite{borgstedt_soothing_2024}.
Haptic heartbeat rendering can foster interpersonal connectedness~\cite{werner2008united},
simulate co-presence~\cite{borgstedt_soothing_2024},
and support emotional self-regulation through mirroring or synchronization~\cite{choi_ambienbeat_2020}.
Heartbeat feedback reinforces a Paro robot's physical and social presence, making it more lifelike and thereby more able to provide socio-emotional support~\cite{borgstedt_soothing_2024}.
We posit that simulated calm heartbeat and breathing behaviors may contribute to an elevated sense of social allyship, and thereby facilitate co-regulation.

We integrated carefully parameterized {breathing} and {heartbeat} 
into the \chora\ used here, for an animate zoomorphic robot capable of influencing users’ bodily states while inviting social connection and eventually, supporting co-regulation. 
These cues should evoke calming responses, first by engaging physiological pathways, 
then adding cognitively 
by  
eliciting the sensed presence of another being, 
much like the comfort one feels when near a calm person or animal~\cite{vyas:2024:happraisalModel}.

\lowHeading{Corroborating past findings}
{We started by confirming a lone past study (Sefidgar~\etal~\cite{sefidgar_design_2016})} which found that even in the absence of instructions on breathing modulation, 
participants who touched a `breathing' robot experienced physiological calming relative to when it was stationary, suggesting that passive perception of a robot's breathing behavior can facilitate ER. 
%
\rev{Corroborating this effect is essential because {aside from it being a sole data point,} the original finding relied on a single robot design, leaving it unclear whether the calming arises from biomimetic animacy alone or its interaction with other design features of that robot platform. 
By testing our zoomorphic prototype, differing in size, shape, weight, appearance, and actuation (Appendix G), we assess whether the effect extends across these two designs, establishing a robust foundation for linking biomimetic animacy to ER facilitation.}
We included~\cite{sefidgar_design_2016}'s behavior parameterizations here \rev{to serve as a reference for our study}.

\subsection{Measuring the Impact of Affective Haptic Interventions}
\label{2:rw:impactmeasure}
A system's ER impact is shaped by an individual's mental health history, context, transient emotion state, regulation style, and emotional susceptibility~\cite{slovak2023designing}. 
Prior work relies on combinations of 
    physiological measures (\eg heart rate, skin conductance~\cite{sefidgar_design_2016}, \rev{cortisol~\cite{sumioka_huggable_2013}, oxytocin~\cite{geva_touching_2020})};
    behavioural proxies (video-coded expressions, body language); 
    touch perception ratings,  system performance metrics, and
    self-report instruments borrowed from psychology (\textsc{sam, stai, erq}), human-robot  (Godspeed) and human-computer interaction (\textsc{nasa tlx, sus})~\cite{vyas_descriptive_2023}.  
However, we lack benchmarks, tools and protocols standardized and validated for affective \textit{haptics} research, and which capture relevant aspects of participant context, including an interaction's subjective experience and its ensuing effects~\cite{vyas_descriptive_2023}, \rev{particularly for ER in this context}.

{Our study adopts a triangulated, mixed-methods~\cite{sefidgar_design_2016}
approach that integrates physiological, self-report, and qualitative data. 
Like others, we contextualized each participant’s background using standardized measures 
    {(\eg relevant demographics and life events, history of depression, anxiety, stress and ER tendencies).}
\rev{As a novel measure,} we deployed a custom affective haptics questionnaire, \rev{developed {with the cognitive psychologist on our team,}} 
grounded in Gross’s ER model~\cite{gross1998:emoRegReview} to assess specific pathways
of \chora's impact.
\rev{To obtain mechanistic 
insights of the \chora\ facilitating ER through participants' perceived experiences, }we incorporated reflective, first-person accounts of felt experience that provided nuanced insights into users' experience, including initial and evolving connection with the \chora, perceived comfort, and relational effects of the touch interaction, to unpack not only \textit{whether} an impact occurred, but \textit{how} and \textit{why} it unfolded within each participant’s emotional landscape.

Together, our measures constitute a step toward a more comprehensive evaluation framework for affective haptics, integrating physiological and subjective measures and accounting for the context-sensitive nature of emotion regulation to enable improved insight into complex regulatory pathways. 

\section{Chora Prototype and Research Testbed}
\label{3:setup}

{We built a \chora\  to evaluate the theoretical model proposed in~\cite{vyas:2024:happraisalModel}. 
This section describes key design decisions, hardware and software implementation and parameterization.}

\begin{figure}[htb!]
    \centering
    \includegraphics[width=\columnwidth]{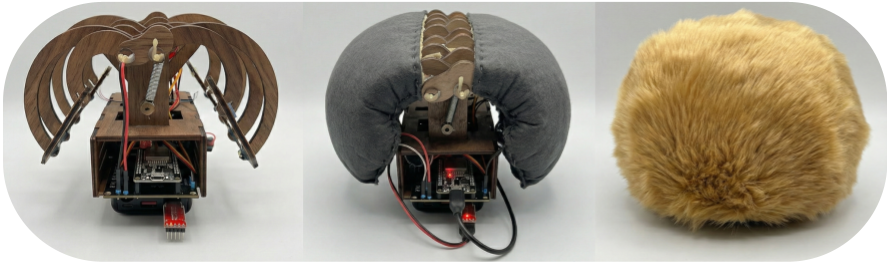}
    \caption{\rev{The \chora\ prototype. From left to right: the instrumented `Ribbit' skeleton~\cite{cang_cuddlebits_2015}; the foam-covered, touch sensor instrumented rib-cage; and the faux fur encasing.}}
    \label{fig:chora-prototype}
     \vspace{-15pt}
\end{figure}

\subsection{\chora\ Prototype}
\label{3:setup:prototype}

{Based on past work, \S\ref{2:rw:behavior}, this} \chora\  is a soft, furry, untethered and roughly spherical zoomorphic robot (largest dimension of \SI{24}{cm})
with parameterized looped behaviors (breathing and heartbeat;
Figure~\ref{fig:s3:researchTestbed}).
Its skeleton is evolved from the wooden ribcage of a `Ribbit' type CuddleBit~\cite{cang_cuddlebits_2015}.

\lowHeading{Shell} 
    The original Ribbit was scaled by 1.5× for manual sensory access to its movement. 
    Faux fur, for comfort and evocation of a small furry animal, topped a 1.5 cm foam layer.
    Added weight provided a lifelike heft (total \SI{800}{\gram}). 

\lowHeading{Movement}
    The breathing mechanism was refined for range, robustness, and precision,  using a \textsc{HS-485HB} servo.
    We added {heartbeat} as a second biomimetic haptic stimulus, with a Titan Haptics DRAKE LF actuator embedded in soft foam within the left ribcage controlled by \textsc{DRV2605} driver. 

\lowHeading{Electronics \& Sensing}
    The base was revised to house a custom PCB, microcontroller, and battery.
    An \textsc{ESP32} microcontroller, connected to the central server via Wi-Fi, coordinated actuation (transmitting server commands) and sensing of user interaction dynamics 
        (IMU, Hall effect, and two different custom touch sensors layered between foam and fur).
        Further sensor and sampling details are omitted as their data is not used in the present analysis. 
    {The entire system} 
    was powered by an on-body \SI{12}{\volt} Charmast \SI{10000}{\milli\ampere\hour} battery and \textsc{CP2102} convertor.

\subsection{{Movement Parameter Optimization Pilot}}
\label{3:setup:movement}


We conducted a parameter setting study (N=10; 5F 5M; ages 21–30) to identify optimized (\OPT) breathing and heartbeat parameters for the \chora\ to be perceived as emotionally comforting and supportive during stressful situations. 
All participants liked animals; seven owned pets. 

\lowHeading{Method}
Participants held a \chora\ 
while adjusting its behavior parameters through a custom interface that updated them in real-time.
\textit{Breathing} {had five tuneable parameters:} amplitude, inhale and exhale speeds, and durations of their preceding pauses; {beats per min (BPM) was derived.}
\textit{Heartbeat}  had three:  `lub' {and} `dub' intensities, and BPM.

Participants explored fixed presets (three comforting, three non-comforting), then fine-tuned them or started from scratch.
They were asked to generate up to three preferred parameter sets (with breathing, heartbeat or both) each for 
holding the \chora\ (a) near their core, or (b) at arm’s length (extremity). 

\lowHeading{Results}
There was no significant difference (paired sample t-test, p$>$0.05) between core and extremity settings for any
parameters. 
Most (9/10) preferred combined \textit{heartbeat} and \textit{breathing}. 
Requests included gentler heartbeat intensity than was available (6/10), and larger breathing amplitude (3).

We identified a \textit{optimized} set of parameters: average values for low-variance parameters, and a range when variance was high.
It featured a
    \textit{heartbeat} that was low-intensity and slow (35–48 BPM)
    with average lub/dub intensity ratio of 1.5;
{combined with} 
    \textit{breathing} within  9–11 BPM,
    {maximal amplitude} 
    with {symmetric} 
    inhale/exhale durations and 
    pause lengths
    with an average {breathing to heartbeat ratio} of 4.0.

\lowHeading{Finalization} 
Following a hardware iteration incorporating the requests above,
we fine-tuned the {pilot-preferred} set,
preserving its core parametric relationships and stimuli preferences, 
    to produce the final \OPT\ condition settings reported in Table~\ref{tab:s4:robot_conditions}. 
    We increased inhale/exhale duration ratio to 3:5 to facilitate calming, reflecting evidence that longer exhale in humans enhances parasympathetic activity~\cite{komori2018relaxation}.
    The final waveforms are detailed in Appendix B.


\begin{table}[htb]
\vspace{-5pt}
\centering
\caption{Experiment conditions and behavior parameters.}
\label{tab:s4:robot_conditions}
\renewcommand{\arraystretch}{1.2}
\begin{tabularx}{\columnwidth}{@{}p{0.55in}@{}p{2.9in}@{}}
\toprule
\textbf{Condition} & \textbf{Description}
\\ 
\hline
\textbf{\ABS} \newline \textit{absent} 
    & No physical contact with or view of the robot, which is placed, turned off, out of sight to participant’s right.  
    \textit{Analysis baseline}. 
    \vspace{1.5pt}\\

\textbf{\INA} \newline \textit{inanimate} 
    & Robot is present but turned off. 
    \textit{Isolates effect of held robot's form, texture, and weight from its movement}.
    \vspace{1.5pt}\\

\textbf{\REF} \newline \textit{reference}
    & Robot moves (breathing) at \textbf{20 bpm, 1.5 cm linear rib movement}, following~\cite{sefidgar_design_2016}. 
    \textit{Reference animate robot condition.}
    \vspace{1.5pt}\\

\textbf{\OPT} \newline \textit{optimized} 
    & Robot moves with optimized breathing at \textbf{9 bpm, 1.5 cm linear rib movement} and heartbeat \textbf{36 bpm, lub 1.5 \micro s, dub 1 \micro s}, parameters tuned in pilot. 
    \textit{Optimized animate robot condition.}
\vspace{4pt}
\\
\multicolumn{2}{@{}p{\linewidth}@{}}
{{In conditions 2-4, the participant holds robot on their lap with non-dominant hand underneath, and gently strokes it with dominant hand (Figure~\ref{fig:s4:studySetup}).}} \\
\bottomrule
\end{tabularx}
 \vspace{-10pt}
\end{table}

\subsection{{Complete Technical Apparatus}} 
\label{3:setup:testbed}


The full setup comprised 
    the \chora, its onboard sensors (not used here), 
    a custom physiological suite, 
     a laptop running a {custom server 
    and control/sampling applications  via the subsystem  microprocessors, {and an LCD display} (Figure~\ref{fig:s3:researchTestbed}).}

\lowHeading{Physiological Sensor Suite}
{An Arduino Uno sampled 
    galvanic skin response (\GSR; Seeed Studio Grove), 
    heartrate (\HR; PPG PulseSensor), 
    and respiration rate (\RR;  Adafruit Conductive Rubber Cord Stretch Sensor)  
    at \SI{25}{\hertz},  uploading them in realtime to the server via a tethered serial link.}

\lowHeading{Control/Collection Software}
A custom application coordinated
    real-time \chora\ actuation,
    experiment condition randomization, 
    timed stimuli and synchronized sampling, 
and 
    graphically displayed live data streams for quality control.

\lowHeading{LCD Instruction Display}
In piloting, we noticed that some participants visually fixated on the robot and missed the slide change. 
Avoiding audio cues to preserve the calm environment, we adjusted slide visual saliency to 26pt Arial font and alternated background colors to highlight protocol transitions.

\section{Methods}
\label{4:methods}
To examine the \chora’s potential to support ER,
we conducted a mixed-methods in-lab user study (N=30).

Many details follow~\cite{sefidgar_design_2016} to allow for comparison: this study is within-subjects, repeated-measures, and manipulates robot behavior as its {sole} factor. 
Departures from~\cite{sefidgar_design_2016}'s protocol include using two parameterizations of animacy (\REF, \OPT), three experimental blocks with interleaved breaks, and administering the demographic survey pre-session.

\subsection{{Conditions, Measures} \& Hypotheses} 
\label{4:meth:cond_hyp}

\lowHeading{Conditions} 
We used four experiment conditions (Table~\ref{tab:s4:robot_conditions}), including two controls and two parameterizations of animacy (as compared to one in~\cite{sefidgar_design_2016}).

\begin{table*}[bt]
\centering
\caption{Overview of data collected. Custom scales are listed in full in Appendix A
{and standard scales detailed in~\cite{vyas_descriptive_2023}.
}}
\label{tab:s4:dataMeasures} 
\renewcommand{\arraystretch}{1.2}
\begin{tabularx}{\textwidth}{@{}p{0.6in}@{}p{6.55in}@{}}
\toprule
\textbf{Phase} & \textbf{Measure Description} \\
%
\toprule 
\textbf{Pre-Study}& 
\textbf{Demographics:} 
    \textit{(1) General items (\textbf{\qDEMO})} (Age, gender, cultural background, education, lived experience, dominant hand, pet ownership, comfort object use); 
    \textit{(2) Affective Object Orientation Questionnaire (\textbf{\qAOO}; custom)} (to capture individual differences in fidgeting behavior, and orientations toward personal objects, including emotional attachment and design sensitivity). 
    \tabrowskip\newline 
\textbf{Trait:} Individual differences in emotion regulation and affective traits using the 
    \textit{(1) Big Five Personality Inventory (\textbf{\qBFI)}}~\cite{john_big_1991}, 
    \textit{(2) Emotion Regulation Questionnaire (\textbf{\qER})}~\cite{gross_individual_2003}, 
    and 
    \textit{(3) Depression Anxiety Stress Scale (\textbf{\qDASS})}~\cite{lovibond_depression_1995}.
    \tabrowskip\\

\textbf{During Blocks}& 

\textbf{Formative Affective State:} Self-reported momentary emotional shifts from 
    \textit{(1) Self-Assessment Manikin (\textbf{\qSAM})}~\cite{hodes_individual_1985} 
    and 
    \textit{(2) State-Trait Anxiety Inventory (\textbf{\qSTAI})}~\cite{spielberger_state-trait_1983}. 
    {Collected 7 times ($\sim$60s each): 
        after all \ABS\ trials including the practice round (4x), and during the 3rd block only for \INA, \OPT\ and \REF\ trials (3x)}.
    \tabrowskip\newline
\textbf{Physiology Stream:} Autonomic arousal related to comfort/stress from 
    \textit{galvanic skin response (\textbf{\GSR})}, 
    \textit{heart  (\textbf{\HR})} and 
    \textit{respiration rates (\textbf{\RR})}.  
    \tabrowskip\newline
\textbf{Behavioral Stream \textit{[excluded from present analysis]}:}  Touch pressure, gesture, speed, duration, and frequency from robot-mounted 
    \textit{flexible \textit{shear sensors}}~\cite{mclaren_what_2024}, 
    \textit{pressure-sensing resistive fabric (EeonTex)}, 
    and 
    \textit{accelerometer}. 
    Nonverbal emotion indicators (facial expressions, gestures, and body posture and movement) from recorded \textit{video}.
    \tabrowskip\\
    
\textbf{Close}&
\textbf{Summative Impact Assessments:}  
    \textit{(1) Internal Dialogue Reflection (\textbf{\qIDR}, custom)} {(to capture participants' thought polarity and content);}  
    \textit{(2) \chora\ ER Facilitation Questionnaire (\textbf{\qCERF}; custom)} (perceived \chora\ support for ER strategies); and
    \textit{(3) Godspeed (\textbf{\qGODSPEED}, modified from~\cite{bartneck_measurement_2009})}, perceptions of the \chora’s zoomorphism, animacy, likability and intelligence). 
    {Completed for participants' most-preferred study condition (identified).}
    \tabrowskip\newline
\textbf{Retrospective Interview:} Topics included participants’ emotional experiences, meaning-making, and narrative interpretations of the \chora’s behavior, both initial and evolving; 
    how the \chora\  may have facilitated or interfered with emotional states and regulation strategies; 
    envisioned use for a \chora\ in daily life;
    and feedback on the study setup.
    \tabrowskip\\
\bottomrule
\end{tabularx}
 \vspace{-15pt}
\end{table*} 

\lowHeading{Measures}
Table~\ref{tab:s4:dataMeasures} details how we assessed participants' affective state via subjective, physiological, and behavioral measures, for insight into ER facilitation.
%

\lowHeading{Hypotheses \rev{(Quantitative)}} \rev{Based on our theoretical framework~\cite{vyas:2024:happraisalModel} and prior observations of haptic calming effects~\cite{sefidgar_design_2016}, we propose the following hypotheses:}


\begin{compactdesc}
 \item  \rev{[$H_1$] 
 \textit{General Calming Efficacy:} 
 Touch interaction with the animate robot (\OPT, \REF) will lead to greater down-regulation compared to 
 the control condition (\INA)
 indicated by $\uparrow$ subjective valence (\qSAMV) and $\downarrow$ subjective arousal (\qSAMA), state anxiety (\qSTAI), and physiological arousal (\HR, \GSR).} 

\item \rev{[$H_2$] 
\textit{Enhancement of Agency:} 
Interaction with an animate (\OPT, \REF) robot will lead to higher perceived dominance (\qSAMD) compared to 
\INA.}

\item \rev{[$H_3$] 
\textit{Additive Effect of Heartbeat:}
Down-regulation extent in the \OPT\ (breathing + heartbeat) will be comparable or greater than that in the \REF\ condition (breathing only).
}
\end{compactdesc}

\rev{{As rationale,} our theoretical framework~\cite{vyas:2024:happraisalModel} proposes that a \chora\ can engage multiple ER processes: \textit{response modulation}, where soothing rhythmic cues directly reduce physiological and subjective arousal 
and increase valence; 
\textit{attention deployment}, as steady, rhythmic haptic patterns redirect attention away from 
perceived stressors, lowering arousal; 
\textit{situation modification}, where the robot functions as an animate, co-regulating partner that enhances perceived control/dominance;
and \textit{cognitive change}, in which biomimetic cues evoke alternative appraisals that elevate valence 
and perceived control. 
}

\rev{Sefidgar \etal~\cite{sefidgar_design_2016} observed calming effects 
($\downarrow$\HR, $\downarrow$\RR, and $\downarrow$\qSTAI; $\uparrow$\qSAMV), 
but found no significant changes in \qSAMD\ or \qSAMA, although both trended upward in active robot trials; they therefore called for verification with additional samples.
Drawing on Self-Determination Theory~\cite{alberts2024designing}, we expected that successful calming would enhance perceived competence, agency, and control over situations~\cite{laban2025robot}, leading to increased dominance.
We further expected the addition of a rhythmic heartbeat in \OPT\ to reinforce the robot’s biomimetic animacy, 
and at minimum yield down-regulation outcomes that remain robust and comparable to those observed in \REF, with the potential to foster a more prominent shift.}

\lowHeading{\rev{Qualitative}}
We anticipated that participants would describe their interactions with the \chora\ as emotionally comforting and conducive to forming a sense of bond or connection.
Based on our pilot, we expected interpersonal variability in preferences between \OPT\ and \REF;
as well as in interaction expectations, employed strategies for facilitated regulation, design preferences, and envisioned long-term integration of the \chora\ into their daily routines.
Given the looped robot behavior, we also expected participants to note feelings of repetitiveness or boredom and a need for interactivity.

\subsection{Experiment Design}
\label{4:meth:design}

\lowHeading{{Sample Size}} 
We targeted a sample of 30 \textit{a priori} using the \texttt{pwrss} online power analysis tool. 
%
We assumed a partial eta-squared ($\eta^2$) of 0.40 from~\cite{sefidgar_design_2016}, 
a Type I error rate of $\alpha=0.05$, a statistical power of $(1-\beta)=0.95$, and four within-subject conditions. 

\lowHeading{Trials}
A single regular trial consisted of a 60s neutralization task (viewing a rippling water video to minimize between-trial carryover) followed by execution of a single condition (75s).
In the initial practice block, both neutralization and trials were shortened to 30s.
These durations followed~\cite{sefidgar_design_2016}.

\lowHeading{Blocks \& Breaks}
Like \cite{sefidgar_design_2016}, each block had 4 trials
{(one per condition)}.
We used three rather than~\cite{sefidgar_design_2016}'s one block to build trust and capture response changes over time (Table~\ref{tab:s4:studyTimeline}). 
A practice session preceded the experiment blocks,
and blocks were separated by a 5m break for washout (word search task and optional physical stretching to balance between cognitive engagement and neutrality).

\lowHeading{Randomization}   
In each block, the baseline \ABS\ condition was always first, and the remaining conditions were counterbalanced with a balanced Latin square.
With N=30, each order was assigned to five participants.
Condition order was further varied across the practice and study blocks by circularly shifting the Latin square matrix to shuffle the order for each round. 
See Appendix C
for full ordering details.

    \begin{figure}[ht!]
    \centering
    \includegraphics[width=\columnwidth]{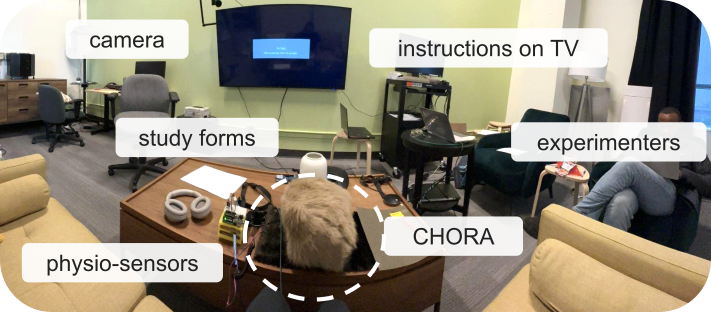}
    \caption{Study setting through a participant's POV.}
    %
    \label{fig:s4:studySetup}
     \vspace{-15pt}
    \end{figure}

\subsection{Study Setup}
\label{4:meth:setup}

For a comfortable, natural environment, we conducted the study in a quiet, dimly lit room with a couch, television, and home-like furnishings (Figure~\ref{fig:s4:studySetup}).
The \chora\ was initially positioned on a soft pad on a low table in front of the participant. 
For right-handers (mirrored for left-handers),  physiological sensors were tethered to the apparatus on the left
of the robot, with forms and a pen on the table.
Instructions were displayed on the TV, which also housed a camera used to record the session via Zoom.

Each session involved two team members. 
The \textit{Lead}
    operated a custom command-line interface to control data collection, including updating the robot mode for each  condition, controlling stimulus duration, logging of touch, physio and video streams to the server, 
    and advancing instruction slides. 
An \textit{Assistant} 
    helped participants wear the physiological sensors,
    and recorded observational notes.
Both experimenters sat to the participant’s right and offered support when needed.


\subsection{Participants}
\label{4:meth:participants}

We collected study data from 30 participants (15F, 14M, 1NB; aged 18–45). 
Recruitment was conducted via a university mailing list and snowball sampling, with a single 
inclusion criteria of age (18+ years).
The sample was demographically diverse (East Asian (12), European (7),  South Asian (5)),
with most holding post-secondary degrees (26). 
Eighteen participants had a pet, and three had also participated in our pilot. 
See Appendix D
for additional details.

\subsection{Procedure}
\label{4:meth:procedure}

Table~\ref{tab:s4:studyTimeline} lists the $\sim$70m session timeline. 
Participants were compensated CAD\$25 for their time, including 
completing a questionnaire and consent form online before arrival.

\begin{table}[t]
\centering
\caption{Study procedure and timeline.}
\label{tab:s4:studyTimeline}

\renewcommand{\arraystretch}{1.2}
%
%
\begin{tabularx}{\columnwidth}{@{}p{0.7in}@{}p{0.35in}@{}p{2.3in}@{}}
\toprule
\textbf{Step} & \textbf{Time} & \textbf{Description} \\
\toprule
\textbf{Pre-Study} & 5 & 
    Demographic and Trait questionnaires.
    \\ \midrule
\textbf{Introduction} & \textit{7} 
    & Consent, overview, instructions, \chora\ introduction, physiological sensors setup, data quality check
    \\ \midrule
\textbf{Practice} & 5 
    & Short version of experiment block\\ 
\textbf{Blocks \textit{(x3)}}
    & \textit{5} & Break (word search and stretch) \\ 
    & \textit{3.25} & Neutralization + \ABS\ + self-report \\
    & \textit{2.25}  & Neutralization + C1 \\
    & \textit{2.25}  & Neutralization + C2 \\
    & \textit{2.25}  & Neutralization + C3 \\
\textbf{Self-report} & 3 & In final block, test conditions C1-3 were also followed by a 1m self-report
    \\ \midrule
\textbf{Close} & 5 & Summative Impact Assessments \\
 & 5 & Retrospective Interview 
    \tabrowskip\\
\hline
 & \textbf{70} & \textbf{minutes {for in-lab portion of study}} \\ 
\bottomrule
\end{tabularx}
\vspace{-15pt}
\end{table} 

\lowHeading{Introduction}
\textit{Lead} confirmed consent, initiated video recording with permission, overviewed the study, including a \chora\ introduction  (without explaining the robot conditions),  explained the self-report questionnaires, and asked the participant to sanitize their hands.

\lowHeading{Sensor Setup}
\textit{Assistant} attached physiological sensors: on  the non-dominant hand, PulseSensor
on index  and \GSR\ electrodes on middle and ring fingers;
a breathing strap around the chest.
While monitoring live signals,  \textit{Lead} asked the participant to take three deep breaths, then lift and re-place the \chora\ to check for comfort and stable signals.
They were asked to avoid rapid movement to reduce signal noise.

\lowHeading{Audio Blocking}
The participant wore noise-cancelling headphones playing pink noise, with volume adjusted to effectively mask ambient sounds and the \chora’s servo motor.

\lowHeading{Instructions}
The participant was asked to watch the TV screen, which displayed the following prompts.
    Neutralization: \textit{``Focus on the video''}.   
    \ABS\ trials: \textit{``Sit still’’}, resting their arms on their lap. \INA, \REF, and \OPT\ trials: \textit{``Hold the \chora\ and interact''} by supporting it with their non-dominant arm and gently stroking it with their dominant hand, as if comforting a pet (Figure~\ref{fig:s4:studySetup}).   
    After each trial: \textit{``Place the \chora\ on the table’’}.
    The slides also indicated when to complete self-report forms or take a break.
Timing was controlled by the data collection software.
\textit{Lead} progressed the slides accordingly.

\lowHeading{During Blocks}
\textit{Lead} monitored physiological signals to ensure data quality.
If excessive sweat was detected on the fingertips (via saturated \GSR\ signals), sensors were temporarily removed and wiped during the next break to maintain signal quality, avoid over-saturation of the signal, and minimize artifacts.
\textit{Assistant}'s notes included the participant's (unprompted) stroking speed and touch interaction style.

\lowHeading{Study Close}
\textit{Lead} asked the participant to complete the final questionnaires considering their preferred \chora\ setting; then conducted the retrospective interview.
Physiological sensors were sanitized between sessions, and the \chora’s outer fabric was replaced halfway through data collection for hygiene.

\subsection{Data Preprocessing, Statistics \& Thematic Analysis} 
\label{4:meth:preproc}
\label{4:meth:stats}

This paper utilizes the following four data components.
For reasons of scope, we address contextual and individual differences and a full behavioral analysis in a separate work.

\lowHeading{Pre and Post Questionnaires}
We computed scores and descriptive statistics for all responses and  corresponding derived metrics for 
    \qBFI\ categories (Extraversion, Agreeableness, Conscientiousness, Neuroticism, Openness), 
    \qDASS\ levels (Anxiety, Stress, and Depression), 
    \qGODSPEED\ dimensions (Zoomorphism, Animacy, Likability, Perceived Intelligence), 
    and the ERQ (Suppression, Reappraisal).

\lowHeading{Within-Block Self-reports}
\qSTAI\ scores were collected using a 9-point Likert scale (‘Not at all’ to ‘Very much’) instead of the original 4-point format, for sensitivity to anticipated participant variance. 
For consistency with validated scoring metrics and cross-measure comparison, the calculated  score range (6–54) was {linearly rescaled} to the standard \qSTAI\ range (6–24).
\textsc{sam} analysis was conducted with N=29, excluding P10 due to missing ratings. 

\lowHeading{Physiological Data}
\GSR\ and \HR\ features were extracted from the final 60s of each 75s  trial, to minimize contamination from transient responses to preceding stimuli and  ensure signal stability.
Raw \GSR\ values were converted to microSiemens and smoothed using a 4-order low-pass Butterworth filter with a 0.05 Hz cutoff. 
\HR\ was derived from the raw photoplethysmography signal using a peak detection algorithm with fine-tuned parameters. 
\RR\ (breathing strap) was excluded due to excessive noise after pre-processing and cleaning.
For each trial, mean values of \GSR\ and \HR\ were computed and subsequently averaged across the three  blocks for a single representative value per condition per participant for statistical analysis.
P19’s \HR\ data was excluded due to excessive noise.

\lowHeading{{Offset Computation}}
To account for individual variability~\cite{sefidgar_design_2016}, 
we \textit{offset} each experimental condition's values (\REF, \INA, \OPT) by subtracting the participant's baseline (\ABS):
\begin{equation*}
\Delta_{\text{cond}} = X^{\text{\textsc{abs}}}_{\textsc{b1}} - \mu_{\text{cond}}, \quad \text{where } \mu_{\text{cond}} = \frac{1}{3} \sum_{i=1}^{3} X^{\text{cond}}_{i}
\end{equation*}
%
Here, $X^{\ABS}_{\textsc{b}_{1}}$ denotes the value from the \ABS\ trial from the first block $(\textsc{b}_{1})$ used as a baseline, and $\mu_{\text{cond}}$ represents the average value of the given condition across the three blocks.
For self-report measures, which were only collected in the third block, $\mu_{\text{cond}}$ corresponds to the single value from that block.

\lowHeading{{Statistics}}
We conducted inferential statistics comparing experimental conditions (\INA, \OPT, \REF) on physiology and self-report measures \HR, \GSR, \qSTAI, \qSAMV, \qSAMA, and \qSAMD.
We assessed \textit{normality} (Shapiro-Wilk) for all six metrics
(Appendix E).
Where indicated, we ran \textsc{Anova} with t-tests for parametric data, and Friedman with Wilcoxon tests for non-parametric data 
(Tables
~\ref{tab:s5:quant-main},~\ref{tab:s5:quant-pairwise}). 
Bonferroni corrections were applied to all \rev{statistical analyses and corrected p-values are reported in the table.}
We conducted these analyses in Python using the pandas, pingouin, scipy, and seaborn libraries.

\lowHeading{Interview {Qualitative Analysis}}
Transcripts were cleaned and coded by two researchers. 
{ER facilitation}-relevant responses were deductively coded using 
the theoretical framework~\cite{vyas:2024:happraisalModel}.
\rev{For rigor and theoretical alignment, first, a junior researcher performed the initial coding, mapping participant responses to ER strategies referenced in relation to the \chora's effects; then 
a senior researcher with expertise in haptic ER theory subsequently reviewed and refined these mapping, with both researchers iteratively clarifying category boundaries to ensure accurate and theory-consistent interpretation.}

The remaining data were inductively coded using reflexive thematic analysis~\cite{braun2019reflecting}, 
\rev{adopting a \rev{collaborative coding approach~\cite{coulston2025collaborative}}.
The two researchers established reliability through iterative dialogue, developing, contesting, and refining codes until consensus was reached on the shared meaning of the themes.}
    
The aspects of this thematic analysis 
relating to {\chora's ER facilitation}
are reported in \S\ref{5:res:movpref}
and \ref{5:res:evolution}. Those relating to design implications and usage requirements are reported separately in a separate publication.

\section{Results}
\label{5:results}

    \begin{figure*}[ht!]
    \centering
    \includegraphics[width=\textwidth]{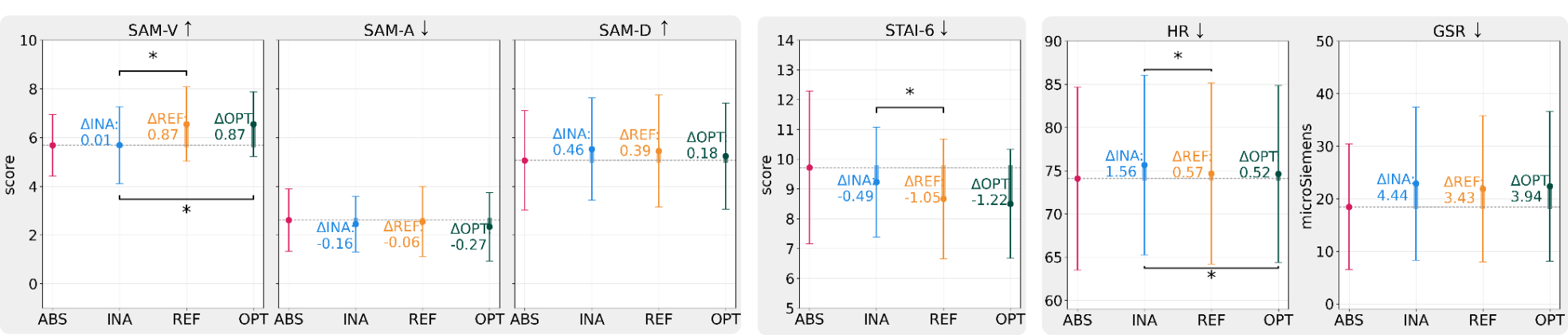}
    \caption{{Quantitative measures for four conditions. 
    $\Delta$ indicates the mean of the respective condition (\INA, \REF\ or \OPT) subtracted from the baseline (\ABS).
    \textit{Annotations:} mean (dot marker), standard deviation (whisker), \ABS\ mean (horizontal dashed line), and offset $\Delta$ from \ABS\ (highlighted vertical lines). 
   {The $\uparrow$ $\downarrow$ in the chart name indicates that metric's hypothesized shift from \ABS.} 
    Statistics are calculated for $\Delta$ values; * indicates metric pairs with p$<$0.05.
    }
    }  
    \label{fig:quant-metrics}
    \vspace{-10pt}
    \end{figure*}

\subsection{\rev{General Impressions of the \chora}}
\label{5:res:gen opinions}

\lowHeading{Self-Reported \chora\ Impact} 
25/30 participants agreed or strongly agreed that the \chora\ \rev{across all conditions} had a positive impact on them, while 2 were neutral, 2 disagreed, and 1 strongly disagreed. 
\rev{We discuss specific preference for \INA, \REF, and \OPT\ conditions in \S\ref{5:res:movpref}.}

\lowHeading{\qGODSPEED}
Participants rated the \chora\ as highly \textit{likeable} (M = 3.99, SD = 0.53), with moderate perceptions of \textit{zoomorphism} (M = 2.97, SD = 0.89), \textit{perceived intelligence }(M = 2.93, SD = 0.68), and \textit{animacy} (M = 2.82, SD = 0.82).

\subsection{\rev{Internal Dialogue and Valence Shift During Experiment}}
\label{5:res:thoughts}

Responses to the \textit{Internal Dialogue Reflection Questionnaire (\qIDR)} indicate an overall positive valence shift of participants’ thoughts over the study (Figure~\ref{fig:s5:customLikertResults}). 
This study did not include a stressor; thoughts' valence distribution was generally neutral to start (20), remained neutral (14) or shifted from neutral to positive (14).
In terms of topics, participants reported thinking mostly about the experiment, themselves (\ie own thoughts, feelings and bodily sensations), and the present moment, 
followed by past and future relationships.
Two participants reported having no specific thoughts throughout. 
Overall, this suggests that participants were largely engaged with the task and aware of their internal states.

\subsection{\rev{Examination of Carry-over Effects}}
\label{5:res:carry over}
\rev{We examined potential carry-over effects across three experimental rounds for \GSR\ and \HR\ (See Appendix F for plots). 
Statistical analyses showed that $\Delta_{\text{cond}}$ \HR\ did not systematically change across rounds (round: $F(2,56)=0.63$, $p=0.54$; condition$\times$round: $F(4,112)=0.23$, $p=0.92$; \textit{Friedman} $\chi^2(2)=0.99$, $p=0.61$). 
$\Delta_{\text{cond}}$ \GSR\ exhibited slight increase over time (round: $F(2,58)=6.06$, $p=0.004$; \textit{Friedman} $\chi^2(2)=6.67$, $p=0.036$), likely reflecting habituation, but with no condition$\times$round interaction ($F(4,116)=0.40$, $p=0.81$). 
Results indicate that any carry-over effects were consistent across conditions, supporting the validity of within-subject comparison of responses across \INA, \REF, and \OPT.}

\begin{table}[h]
\centering
\caption{Inferential statistics for $\Delta$ metrics. p$<$0.05 in \textbf{bold}.}
\label{tab:s5:quant-main}
\begin{tabular}{@{}p{1.6cm}p{1.8cm}p{1.4cm}p{0.4cm}r@{}}
\toprule
\textbf{Measure (n)} & \textbf{Test Type} & \textbf{Statistic} & \textbf{\rev{Corr. p-value}} & \textbf{Effect Size} \\
\midrule
\qSAMV (29)   & Friedman (2)     & $\chi^2$ = 12.79    & \textbf{0.001} & W = 0.22 \\
\qSAMA (29)   & Friedman (2)     & $\chi^2$ = 0.83     & 0.660 & W = 0.01 \\
\qSAMD (29)   & Friedman (2)     & $\chi^2$ = 0.84     & 0.657 & W = 0.01 \\
\qSTAI (30)  & ANOVA (2, 58)    & $F$ = 6.46          & \textbf{0.003} & $\eta^2$ = 0.02 \\
\HR (29)      & ANOVA (2, 56)    & $F$ = 5.49          & \textbf{0.007} & $\eta^2$ = 0.03 \\
\GSR (30)     & Friedman (2)     & $\chi^2$ = 4.07     & 0.13 & W = 0.07 \\
\bottomrule
\end{tabular}
\vspace{-10pt}
\end{table}
\begin{table}[ht]
\centering
\caption{Summary of post-hoc pairwise tests.}
\label{tab:s5:quant-pairwise}
\begin{tabular}{@{}lccc@{}}
\toprule
\multicolumn{1}{l}{} & \multicolumn{1}{c}{\textbf{\qSAMV}} & \multicolumn{1}{c}{\textbf{\qSTAI}} 
& \multicolumn{1}{c}{\textbf{\HR}} \\
\midrule
{$\Delta_\textsc{INA}-\Delta_\textsc{REF}$}  &  Wilcoxon & T-tests & T-tests \\
\midrule
mean    & -0.86   & 0.73  & 1.03    \\

std & 1.19   & 0.98   & 2.24    \vspace{4pt} \\

Test Statistics (W/t)             & 0.0   & 4.10 & 3.34    \\

Corr. p-value                 & \textbf{0.002} & \textbf{0.0009} & \textbf{0.007}  \\

Effect Size ($r^2$)           & \textbf{0.780} & \textbf{0.367} & \textbf{0.285}  \\

\midrule

{$\Delta_\textsc{INA}-\Delta_\textsc{OPT}$} &  &  &  \\
\midrule
mean    & -0.86   & 0.57  & 0.99    \\

std & 1.51   & 1.36   & 1.59      \vspace{4pt} \\ 

Test Statistics (W/t)                    & 18.0   & 2.29 & 2.49    \\

Corr. p-value                  & \textbf{0.004} & 0.09  & \textbf{0.019}  \\

Effect Size ($r^2$)           & \textbf{0.451} & \textbf{0.153} & \textbf{0.181} \\

\midrule 
{$\Delta_\textsc{OPT}-\Delta_\textsc{REF}$} &  &  &  \\
\midrule
mean    & 0  & 0.17 & -0.05    \\

std & 0   & 1.15   & 1.81      \vspace{4pt} \\ 

Test Statistics (W/t)                 & 59.0   & 0.80 & 0.14    \\

Corr. p-value                  & 0.98 & 1  & 0.89 \\

Effect Size ($r^2$)          & 0 & 0.021 & 0.001 \\
\bottomrule\\
\multicolumn{4}{r}{*$p < 0.05$ and $r^2 > 0.14$ values in \textbf{bold}}
\end{tabular}
\vspace{-5pt}
\end{table}

\subsection{\rev{Quantitative }Evidence of the Animate \chora's 
Calming}
\label{5:res:passiveCalm}

\begin{table}[t]
\centering
\caption{\rev{Summary of hypotheses and their outcomes}.}
\label{tab:hypotheses}
\begin{tabular}{p{0.69\linewidth} p{0.188\linewidth}}
\toprule
\rev{\textbf{Hypothesis}} & \rev{\textbf{Outcome}} \\
\midrule
\rev{\textbf{$H_1$}: Down-regulation extent will be greater in the animate conditions (\OPT, \REF) than in the control condition (\INA).}
& \rev{Supported}
\\ 
\addlinespace
\rev{\textbf{$H_2$}: Animate conditions (\OPT, \REF) will lead to higher perceived dominance compared to \INA.}
& \rev{Not Supported}
\\
\addlinespace
\rev{\textbf{$H_3$}: \OPT\ will elicit comparable or stronger down-regulation than \REF.}
& \rev{Supported (comparable)}
\\
\bottomrule
\end{tabular}
\vspace{-12pt}
\end{table}

\rev{We assessed quantitative measures 
for the \chora's 
\textit{passive} (\ie without participant intent or awareness) 
impact on physiology and affective state.
}

Figure~\ref{fig:quant-metrics} shows descriptive statistics for absolute 
and mean delta values (further details in Appendix E);
Tables \ref{tab:s5:quant-main} and \ref{tab:s5:quant-pairwise} list inferential statistics comparing \dINA, \dREF\ and \dOPT\ by metric, \rev{and Table~\ref{tab:hypotheses} hypothesis outcomes.}
In summary, 3 of 6 measures (\qSAMV, \qSTAI, \HR) revealed statistically significant differences between \dINA\ and \dREF, and two  (\qSAMV, \HR) differentiated \dINA\ and \dOPT.
We found no significant differences between \dOPT\ and \dREF, although the latter had larger effect sizes. 
\textbf{These findings corroborate \cite{sefidgar_design_2016}}, where interactions with an animate robot led to an increase in valence, decrease in state anxiety, and reduction in heart rate.

\subsection{\rev{Self-reported}
Preferences of Different \rev{Conditions}}
\label{5:res:movpref}


Although not explicitly informed about robot conditions, 
participants were asked in the 
post-interview whether they noticed differences in the \chora's behavior within blocks and, if so, which they preferred.
\rev{Based on their description, we matched each response to the relevant 
study condition.}

\lowHeading{Noticing} 
All participants differentiated \INA\ from animate conditions (\OPT, \REF).
20 participants noticed clear difference between \OPT\ and \REF,  4 a subtle difference or were uncertain; 6/30 did not notice any change.
Three distinct behaviors were described: 
    a slow breathing motion (22/30), 
    a rhythmic heartbeat (18), 
    and a soft purring-like sensation (4).
The purring was unintended,  a vibration of the servo motor that generated the breathing motion.

\lowHeading{Preferences}
29/30 preferred animate conditions (\OPT, \REF) over \INA, noting that movement enhanced the \chora’s perceived aliveness and emotional resonance.
Conversely, one felt ``\textit{more in control}'' when the robot remained still (P19).
{Breathing salience seemed desirable.}

\rev{3}/29 preferred \REF\ 
because breathing was more haptically perceptible, possibly because of \REF's higher breathing rate. 
\rev{Among these, two showed $\uparrow$ \qSAMV, $\downarrow$ \qSTAI, and $\uparrow$ \HR\ in \REF, whereas the other showed no change in \qSAMV\ or \qSTAI\ but $\downarrow$ \HR\ in \REF\ compared to \OPT.}

\rev{13}/29 participants primarily \rev{preferred the heartbeat stimuli for its `calming',  `life-like' effect.}
\rev{6/13  
    favored \OPT's rhythmic heartbeat
    combined with its lower breathing rate, which came across as softer.} 
%
%
%
%
\rev{Among the participants who articulated heartbeat's calming effects, 
7 showed $\downarrow$ \HR\ in \OPT\ relative to \REF\ ($\mu = 1.95$, $\sigma = 1.45$), whereas 2 showed no difference and 4 showed $\uparrow$ \HR\ ($\mu = 1.25$, $\sigma = 0.32$). 
Self-reports showed limited alignment with these physiological changes, with the majority reporting stability: for \qSAMV\ ($\uparrow$ 5/13, $\downarrow$ 1/13, same 7/13) and \qSTAI\ ($\downarrow$ 4/13, $\uparrow$ 1/13, same 8/13). 
Thus, while the presence of heartbeat (in \OPT) was associated with higher calming effects than \REF\ for a subgroup of participants, the overall pattern indicates meaningful individual variability.}

\rev{3/29 specifically mentioned preference for purring-like movement, which was an artifact of the servo motor in both \REF\ and \OPT\ conditions; whereas the remaining 10/29 preferred animate movements but were not able to perceptually distinguish between \REF\ and \OPT.} 
        




\subsection{ \rev{Participant-{Reported}} 
ER Facilitation}  
\label{5:res:paths}

In the retrospective interview, we asked participants to describe their emotional experiences throughout the session, including how they felt, how their emotions changed over time, any ER strategies that they invoked, and noting whether and how the \chora\ played a role.
Interview data is supplemented with responses to the \textit{\chora\ ER Facilitation Questionnaire} (\qCERF), which address both strategies used and perceived impact (Figure~\ref{fig:s5:customLikertResults}).
This section is organized around the four haptically facilitated ER strategies identified in \S\ref{2:rw:chora}.

\rev{Although interview responses were not segmented by experimental condition, the analysis emphasizes the specific \chora\ features that participants reported as influencing their emotional experience. 
This framing situates the qualitative findings in design-relevant elements, such as form, appearance, user interaction, and robot behavior, and the mechanisms 
through which these features facilitated ER, rather than focusing on a condition-by-condition comparison, which provides less insight into the cross-cutting features driving physiological and emotional responses. 
References to movements, breathing, and heartbeat can be interpreted in context: heartbeat corresponds primarily to the \OPT\ condition, breathing to \REF\ and \OPT\, and form and appearance are relevant across all conditions.}

    \begin{figure*}[ht!]
    \centering
    \includegraphics[width=1.0\textwidth]{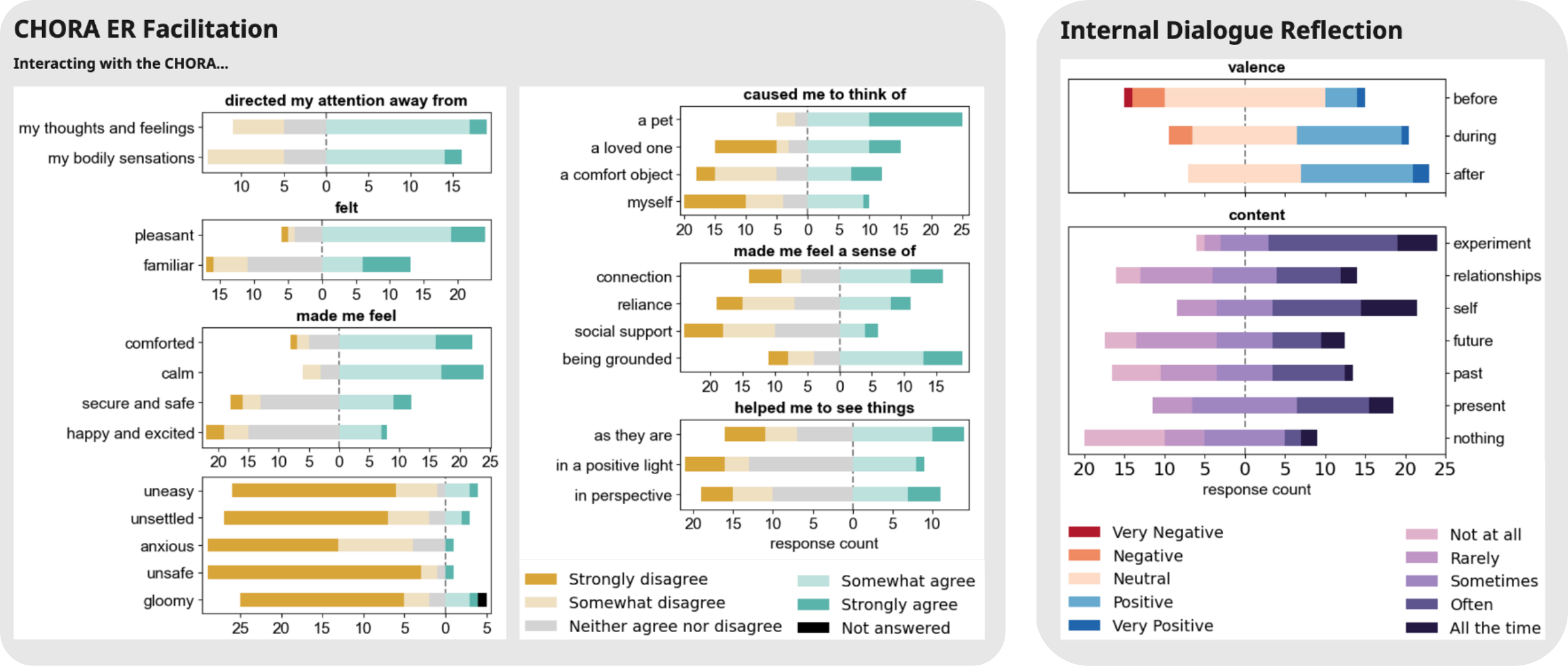}
    \caption{Responses from \textit{Internal Dialogue Reflection} (\qIDR) and \textit{\chora\ ER Facilitation} (\qCERF) questionnaires.
    \rev{Participants completed these items with respect to their preferred \chora\ conditions; 29/30 preferred animate conditions.}
    }
    \label{fig:s5:customLikertResults}
    \vspace{-15pt}
    \end{figure*}

\subsubsection{\textbf{The \chora\ Promotes Physiological Calming}}
\label{5:res:paths:physio}

77\% of participants (23/30) reported that the \chora,  held to their core and in the condition they preferred, helped soothe their emotions and foster a sense of calm, mentioning its form factor, texture, and movement, and their inclination to stroke it.
This count (23) comprises descriptions that indicated direct emotion down-regulation, rather than distraction (\S\ref{5:res:paths:attention}).

\lowHeading{Quantitative (\qCERF)}
The clear 
    rightward skew on positive items (feeling pleasant, comforted, calm,  grounded) 
    and leftward skew of negative items (uneasy, unsettled) 
suggests that \textit{physiological calming} was key for 83\% (25/30).

\vspace{0.05in}\noindent
Participants mentioned several aspects of calming.

\lowHeading{{Form \& Texture (3/30)}}
The haptic qualities (small-animal size and soft, plush texture) evoked associations with pets or soft toys: 
\textit{``It felt like interacting with a real animal, like a big cat \dots creating a calming experience''} (P18);
\textit{``It reminded me of holding something soft and comforting, which helped me feel more relaxed''} (P10).

\lowHeading{{Heartbeat (\rev{13}/30)}}
%
This movement was \textit{``soothing,'' ``natural,'' ``familiar,'' ``lifelike''} and \textit{``emotionally resonant''}:
    \textit{``The sensation of a heartbeat from \chora\ helped calm me a lot. It was a very good feeling, and it helped me relax to the point that I almost dozed off''} (P1);
    `\textit{`When it had the heartbeat-like movement, I felt the most positive, calm, and peaceful because it felt the most lifelike''} (P17).
    However, {\rev{2}/30} diverged, describing it as \textit{``creepy''} \rev{(P26, P15)} and \textit{``unnatural''} (P26). 
 
    %

\lowHeading{{Breathing (5/30)}}
17\% described the \chora's breathing as ``\textit{natural}'' and ``\textit{calming}''.  However, 3/30 were put off.
Some found it unrealistic:
    \textit{``The lack of visible movement with the breathing created a bit of a disconnect''} (P16);
    \textit{``When it was doing the breathing movement, it was calming but somewhere between positive and neutral.''} (P17, who preferred heartbeat as most lifelike). 
P13 and P23 found the same repetitiveness boring that P21 said contributed to calming.


\lowHeading{{Holder's Actions (4/30)}} 
Soft stroking and holding the robot close to the body (instructed) acted as an amplifier. 
    ``\textit{Steady and calm heartbeat, combined with patting it, helped me follow its rhythm}'' (P5).
Others touched less rhythmically:
P11 mindfully 
stroked to observe changes in the \chora’s behavior and reported calming arising from a perceptual-reflective cycle.
P6 used the \chora's movement to synchronize their breath.

\lowHeading{{Via Attention Deployment (7/30)}}  
{As elaborated below, \chora\ movement captured and focused attention. 
This was specifically identified as contributing to physiological calm.}

\subsubsection{\textbf{The \chora\ Redirects Attention}}
\label{5:res:paths:attention}
40\% of participants (12/30) described attentional shifts in a way that suggested interplay between external engagement and internal cognitive states, 
both \textit{toward} the \chora\ as an animate agent, and \textit{away} from visual stimuli, bodily discomfort, or intrusive thoughts, particularly negative ones. 
They connected the \chora's calming behaviors (breathing, heartbeat) to facilitating and sustaining these shifts.

\lowHeading{Quantitative (\qCERF)}
19/30 participants reported attention shifting away from thoughts; 16/30 noted redirection from bodily sensations.

\lowHeading{{Movement} as a Captured or Intentional Focus (12/30)}
    %
\rev{Participants reported that} the \chora's movement \textit{drew} attention, supplanting other people (P21),  visual stimuli (P27) {or own body (P22)}. 
 It also provided a focus which helped them detach from recurring thoughts and mind wandering:
     ``\textit{While interacting with \chora, I had fewer thoughts coming to my head}'' (P2). 
     ``\textit{it gave me something to focus on, similar to when cats purr; it’s very comforting}'' (P16).
This could be deliberate, used like a ``\textit{tool}'' (P30):  
``\textit{I intentionally tried to direct my focus to the robot ...}'' (P30).  
``\textit{I think it drastically changed when the robot started with the heartbeat and breathing}'' (P5, echoed by P12, P14).

\lowHeading{{Grounding} (6/30)} 
Focus could lead to anchoring outside of one's self:  
    ``\textit{... while I was watching the waves [neutralization], I noticed I was paying attention to my body, but when I was interacting with the \chora, I wasn’t so much focused on my body anymore ... being in the present helped me process things pleasantly}'' (P22).
    \chora's movement helped: ``\textit{When \chora\ was moving, I ... 
    felt connected to its actions}'' (P12).

\lowHeading{{Entitlement of Animacy} (10/30)}
33\% explicitly linked their attention deployment to the \chora’s behaviors, and its implied animacy and agency increase their willingness to engage.
Three felt that its life-likeness warranted their attention:
    the active \chora\ felt ``\textit{more like an animal, more deserving of my attention}''; inactive, it felt like ``\textit{a toy}'' (P14);
    P5 and P12 felt obligated to pay attention when it moved.

\lowHeading{Heartbeat Anchoring (3/30)} 
The heartbeat attracted comment separately from the breathing:
    ``\textit{It was easier to focus}'' when both heartbeat and breathing were present (\OPT), whereas breathing movement alone (\REF) led to more distraction and self-focused thoughts (P6). 
    ``\textit{The steady and calm heartbeat, combined with patting it, helped me follow its rhythm.  I slowed down my motion, which in turn slowed down my thoughts, as I tend to think very fast}'' (P5).

\lowHeading{{Boredom w/o Responsiveness} (1/30)}
This \chora's movement wasn't always enough; interactivity might help (P13).

\subsubsection{\textbf{The \chora\ is Perceived as an Ally}}
\label{5:res:paths:ally}


73\% of participants (22/30) described the \chora\ as a comforting presence or a companion.
They compared it to a pet or even a ``\textit{sleeping infant}''
based on its form (``\textit{fluffy}'',``\textit{cuddly}''),  rhythmic behaviors, attributed personality (``\textit{friendly}'') and its broader context.

\lowHeading{Quantitative (\qCERF)}
Although items related to the \chora’s \textit{functioning as an ally} (fostering connection and reliance) showed a wide distribution,  83\% (25/30)  agreed or strongly agreed with statements likening the \chora\ to a pet, and half (15/30) endorsed viewing it as a loved one.

\lowHeading{{{Perceived }Animacy (20/30)}}  
The simulated calm {heartbeat and breathing} contributed to this connection.
Even without being responsive, these behaviors
evidently shaped the \chora\ as a soothing co-regulator by evoking petting a calm animal and grounding in familiar, embodied experiences of comfort.
13/30 said the  movements made it feel alive
    (``\textit{purring}'', ``\textit{buzzing}'', ``\textit{vibrating}'', ``\textit{breathing}'', having a ``\textit{heartbeat}''); 
    or more explicitly, 
    like ``\textit{a living object}'' (P27)
    or ``\textit{petting a breathing dog}'' (P17, P29).
These physical sensations evoked calming associations, \eg to 
    a ``sleeping animal'' (P28);
    or nearly dozing off due to its rhythmic heartbeat (P1). 
        

Across conditions, 
some attributed meaning to the \chora's changing behaviors.
Movement was interpreted as a sign of vitality, and stillness as illness, fatigue, or death by eight participants, among which five anthropomorphized \chora\ in inactive trials with their concern:
    ``\textit{[it] felt real and alive'' only when it moved}'' (P25); 
    ``\textit{...  it stopped moving. I think I feel a little more worried because I'm sort of treating it as a pet...   so I'm wondering, are you sick or something?}'' (P7); 

This cessation evoked such a strong memory of a pet passing away for P9 that their \GSR\ and \HR\ increased markedly compared to active trials.
Obviously, an individual's sense-making is shaped by their experiences. 
This sample \rev{illustrate an account of}
suspension of disbelief and motivates a closer examination of individual differences.

\lowHeading{{Companionship (15/30)}}
Several highlighted a {perceived bi-directional social bond}. 
They offered care:
    \textit{``...when the robot started with the heartbeat and breathing, I felt like there was another kind of lively object around me, and I should pay attention to it, support it, or transfer positive emotions to it''} (P5);
and received it:
    
    ``\textit{it just kind of like, oh my gosh, like I'm cuddling this fluffy animal, I'm gonna be fine}'' (P30).

P2 felt curious about ``\textit{what \chora\ was doing and how it felt}''.
Some  sensed situational co-presence, saying they were
    ``\textit{watching the (neutralization) video [alongside the \chora]}'' (P2);
    ``\textit{chilling on the sofa, watching Netflix [with a pet]}'' (P29).
This social attribution aligns with theories of perceived social presence~\cite{eckstein_calming_2020, borgstedt_soothing_2024}
and ER through co-regulation.

\lowHeading{{Negative Social Reactions (5/30)}}
Five participants expressed discomfort due to a mismatch between their expectations of animacy and \chora\ movement.
    The heartbeat seemed ``\textit{creepy}'' and ``\textit{unnatural}'', generating unease (P15, P26);
    movements were ``\textit{weak}'' for a living animal (P26);
    while reminiscent of their cat, it needed responsiveness to seem alive (P23, P15).
P19 described a deeper discomfort with the robot:
    ``\textit{It felt a little strange because I know it’s an inanimate object trying to replicate being something animate. That made me feel uncomfortable, to be honest. I almost preferred it when it didn’t move because I felt slightly more in control and grounded in the reality of the situation.}''
Here, cognitive dissonance arose in an inanimate object mimicking life.
These critical perspectives 
underscore a complex interplay between expectations, embodied experience, and cognitive processing.

%

\subsubsection{\textbf{The \chora\  Leads to a Cognitive Shift}}
\label{5:res:paths:cognitive}
43\% of participants (13/30) described how the emotional and sensorial calming evidenced above 
prompted more cognitive activities: reflective thought, resurfacing of memories, or subtle shifts in perspective. 
These accounts offer early evidence that the \chora’s calming effect facilitated cognitive flexibility by helping users disengage from immediate emotional strain and access broader or alternative mental frames.

\lowHeading{Quantitative (\qCERF)}
Responses on questionnaire items associated with \textit{cognitive {change}}
were promising. 
14/30 participants agreed that interacting with \chora\ helped them appreciate things as they are; 11/30 reported gaining perspective, and 9/30 viewed their situation in a more positive light.
However, variability points to important individual differences.

\lowHeading{Reframing of Thoughts (7/30)} 
%
%
While initially preoccupied with stressors like homework and exams, engaging with the \rev{animate} \chora\ helped P25  ``\textit{view plans more positively}'' and 
P8 ``\textit{stabilize thoughts}''; while both positive and negative thoughts arose, they felt more balanced and reflective.
The interaction \rev{with {the} animate \chora} helped P14 and P30 to reframe, to ``look at life differently'' and ``put things in perspective'' \rev{as they perceived it as a real-life animal}.
However, P3 reported no \chora\ impact on their thoughts, which remained focused on work, research, and problem-solving.

\lowHeading{Direct Cognitive Path {via Memory Invocation} (10/30)}
{We found evidence 
of touch interactions expanding positively valenced cognition.}
\rev{Participants reported that} \chora\ interactions evoked memories 
    of pets (7), 
    family relationships (4), and 
    emotionally significant experiences (3): 
    ``\textit{... when I was stroking the \chora\ and feeling the two different movements, I started thinking about my dog ... then I began thinking about my past, specifically good memories}'' (P17);
    ``\textit{... memories of a cat I met while running the other day, which made me feel great}'' (P20).
{However,} P9's pet memory was a rather sad one.

    %
    %
    %
\rev{Participants also described that the \chora}
activated salient memories that were less `used', which could be useful in reappraisal.
    ``\textit{Heartbeat sensation brought up recent family memories, which was interesting because I don’t often think about family. These thoughts might have been related to the comfort aspect ...}'' (P24); 
    ``\textit{I just started missing my baby cousins—they’re not babies anymore—and remembering playing with them}'' (P21).

\subsection{Initial \vs Evolving Reactions to \chora\ Behaviors}
\label{5:res:evolution}

To examine how novelty shaped participants’ impressions, we asked how they felt when the \chora\ first moved and how their reactions might have evolved.
\rev{Two researchers inductively coded the response using collaborating coding~\cite{coulston2025collaborative}, generating}
four themes capturing participants' reflections on their       
    \textit{animacy association} with,
    \textit{behavior sense-making} of, 
    \textit{emotional reaction} to, 
and \textit{perceived impact} of the \chora.

\lowHeading{Initial Reactions}
{9/30} participants {evidently inferred initial \chora\ \textit{animacy} by associating} 
it with a {pet}; {conversely}, two described it as {mechanical or robot-like}. 
7/30 {engaged} 
in early {efforts at} \textit{behavior sense-making}; 
    {\eg  curiously} noticing what changed between each trial (3), 
    {or expressing confusion as to}
    whether the \chora\ was \textit{``responding''} or ``\textit{malfunctioning}'' (4): 
P26 said they were \textit{``holding [their] breath''} to figure out what would happen next.
Some participants reported their first \textit{emotional reaction} as {excited} (8/30) or {surprised} (2), and varying \textit{perceived impacts}, \eg of both {comfort} (2) and {discomfort} (2).
16/30 did not mention any specific initial emotional reaction or perceived impact.

\lowHeading{Evolution}
{Participants identified shifts in both their perception of the \chora\ as pet-like (more animate), and in their familiarity with it.
5/30 ultimately joined the nine who deemed it pet-like from the start:} 
``\textit{By the end of the experiment, interacting with \chora\ felt like second nature, almost like holding a living pet.}'' (P10).

A greater number (15/30) reported feeling {familiar} with the \chora\ and its behaviors over time; \eg
    ``\textit{The motion became less surprising and more predictable as the experiment progressed}.'' (P18).
{In some cases this was expressed as a positive, although not specifically linked to a change in animacy interpretation, \eg}
    ``\textit{Over time, I grew accustomed to the movement, and it became a calming and enjoyable experience.}'' (P1). 

{For some, this led to an enjoyable ability to \textit{anticipate}}
the \chora's behaviors (5/30):
    ``\textit{Later, as I became more familiar, I started to anticipate its movements and felt more natural interacting with it}'' (P8);
{or perhaps conversely, led them into engaged \textit{sense-making}, continually {observing} their interactions} (7/30). 
{However, the movement's repetitive nature led to disengagement for others} (5/30). 

\lowHeading{Summary}
{Individuals interpreted the same stimuli in different ways. 
Initial reactions varied, often shaped by memories and associations, or by novelty; and 
evolution clearly occurred, with both familiarity and sense-making  in play. 
Outcomes ranged from 
    growing attachment and ability to find comfort based on relational interpretations, affinity with the movement, or both; 
    to disengagement due to boredom with repetitive or incoherent behavior.
At minimum, it is evident that novelty is not always just noise, but can catalyze meaning-making.
}

\section{Conclusions and Future Work}
\label{6:conc}
%
This study examined whether and how touch interaction with a zoomorphic \chora\ exhibiting calming bio-mimetic animating behaviors facilitates passive calming, 
{drawing on evidence from self-report, physiological, and qualitative data.}

\subsection{Corroboration \rev{of Past Findings on Passive Calming}} 
\label{6:conc:Aim1corroboration}

\rev{We aimed} to replicate and estimate parametric sensitivity of prior findings~\cite{sefidgar_design_2016}.
Our prototype retained~\cite{sefidgar_design_2016}'s basic idea: a lapsized, a zoomorphic robot,
that breathes and invites touch.
It differed in physical construction details and behavior (added a heartbeat and a different parameter optimization). 

{Like~\cite{sefidgar_design_2016}, we found that} affective touch interactions with an animating \chora\  led to increased subjective valence, decreased physiological arousal, and reduced state-trait anxiety \rev{($H_1$)}. 
This demonstrates the effect's robustness in a modified robot prototype, supporting
the regulatory effect's generalizability across \rev{these two} design variations. 

Both studies (here and~\cite{sefidgar_design_2016}) showed that it was the addition of slow, rhythmic, haptic stimuli to
    a socially evocative, zoomorphic, and touch-inviting form that generated the regulatory effect.
The similarity of responses across varied active conditions reduces the need for sensitivity testing, although it would be valuable to establish outer limits.  
However, we do not yet know whether the heartbeat would work on its own, or (we suspect less likely) an arrhythmic haptic cue, or a less biomimetic one. 
Examining these would further inform the 
{mechanistic origins}: 
What is it about the haptic cues that trigger the response we have observed? 

\subsection{\rev{No Difference in Subjective Arousal and Dominance}}
\rev{Similarly to~\cite{sefidgar_design_2016}, we observed no significant group-level differences in subjective arousal or dominance across conditions ($H_2$). 
Participants’ qualitative accounts suggest heterogeneity in how these dimensions were experienced. 
Some engaged with the \chora\ as an interactive or attentional tool, which might have heightened arousal, whereas others used it primarily for soothing and reflection, which might have lowered arousal. 
Perceptions of dominance similarly varied. 
We reflect that a caregiving orientation toward the \chora\ might have increased users’ sense of control, whereas relying on it as a tool for regulation might have reduced perceived dominance, as participants might have perceived the robot as taking a more active role, with them in a receptive position.}

\subsection{\rev{No Additive Effect but Selective Preference of `Heartbeat'}}
\label{6:conc:heartbeat effect}
\rev{Aggregate physiological measures (\HR, \GSR) and self-reports (\qSTAI, \qSAM) did not show a statistically significant additive effect of the rhythmic heartbeat behavior in \OPT\ compared to \REF, suggesting that, on average, the presence of heartbeat did not produce detectable changes in measures of physiological or affective calming.
These results indicate comparable down-regulation performance between \OPT\ and \REF\ {($H_3$)}.}

\rev{However, qualitative data revealed important nuance. 
13/30 participants expressed a selective preference for the heartbeat, 
noting that it enhanced perceived animacy, connection, and comfort while anchoring attention. 
This indicates that, although cumulative quantitative measures did not show a statistically significant additive effect, the heartbeat successfully enhanced the perceived animacy and relational quality of the \chora\ for a subset of participants. 
These varied subjective responses to the \chora\ reflect contextual and personal preferences for haptic cues, highlighting the need for future work on person- and context-specific tailoring of \chora's behaviors to maximize co-regulatory potential.}



\subsection{{Evidence of \chora-Linked} 
ER Facilitation}
\label{6:conc:Aim2indirectPath}

We sought to validate 
our theoretical framework for haptically-supported ER~\cite{vyas:2024:happraisalModel}.
Through mixed-methods analysis, \rev{we identified several ways in which participants perceived and articulated that touch interaction with the \chora\ facilitated their ER strategies:}

\lowHeadingNoPunct{Promoting physiological calming}
    through form and texture, heartbeat, the holder's actions, and through their attention.  
    
\lowHeadingNoPunct{Redirecting attention}
    by capturing it or offering a focus, external grounding, and animacy-based entitlement to it.   

\lowHeadingNoPunct{Functioning as an ally}
    first by seeming alive, then by seeming a companion. We also  saw negative impacts when the \chora's behavior violated expectations.  

\lowHeadingNoPunct{Leading to a cognitive shift}
    via thought reframing and invocation of memories, including infrequent ones. 

\rev{Both inanimate and animate \chora\ features elicited ER effects; however, animate behaviors of heartbeat and breathing produced more pronounced impacts on physiological calming, attentional focus, perceived companionship, and cognitive reframing, as reported by participants in their self-reports.}
\rev{These findings, derived from participants’ reflective self-reports (\S\ref{5:res:paths}) and complemented by quantitative evidence of passive calming (\S\ref{5:res:passiveCalm}), provide convergent support for our theoretical framework~\cite{vyas:2024:happraisalModel}.}

\subsection{Methodological Reflections and Limitations}
\label{6:conc:limitations}

\lowHeading{Limits to Generalization}
{We demonstrated a regulatory effect for a given \chora\ relational framing and across modest variations in its form and stimulus details.
However, while prior research~\cite{vyas_descriptive_2023} suggests promise, this does not yet extend to} 
other \chora\ embodiments (\eg wearable~\cite{choi_ambienbeat_2020},
humanoid~\cite{block_softness_2019}). 
Our framework~\cite{vyas:2024:happraisalModel}, agnostic to \chora\ form, could help to steer investigation of how 
 and to what degree different \chora\ types could facilitate ER strategies.

\lowHeading{{`As you are' Emotion Baseline}} 
We prioritized observing the effect {on participants' baseline of an animated \chora, unclouded by 
the additional manipulation of} an explicit emotion elicitation task.
While this {meant that participants' baseline emotion states were not calibrated,} 
it also minimized interpretive noise and the risk of overstating effects, particularly relevant given the subtle nature of the haptic intervention. 
We believe this made our test more stringent, \ie muting effects rather than inflating them. 
In subsequent research, we plan to use narrative-based or ecologically valid elicitation methods
to probe how these mechanisms 
function under more naturalistic ER demands.

\lowHeading{\rev{Physiological and Self-Report Measures}}
\rev{We measured participants’ physiological responses using heart rate (\HR) and galvanic skin response (\GSR), alongside self-report scales including the \qSTAI and \qSAM. 
While these metrics provide complementary insights into arousal, valence, and stress regulation, their interpretation warrants caution. 
For \HR, \GSR, and \qSAMA, increases or decreases could both reflect positive engagement, such as playfulness, focused attention, or relaxation, depending on individual differences. 
Similarly, \qSAMD\ responses may vary: the presence of the \chora\ could increase feelings of dominance when participants perceive control over the interaction, or conversely, decrease dominance when the robot appears to take control, eliciting a sense of submission or being ‘smaller’ in the interaction. 
Self-report measures capture subjective experience but are susceptible to bias, demand characteristics, and variability in introspective accuracy. 
Together, these factors highlight the importance of triangulating across multiple data sources and carefully contextualizing physiological and self-reported effects when drawing conclusions about ER facilitation.}

\lowHeading{{Effect of Touch Constraints \& Experiment Setup}} 
{At the study's start,} participants were guided to `gently stroke' the \chora;
{subsequent interactions were monitored but not corrected.
This instruction targeted construct validity, through a relatively consistent touch experience; and reinforcement of the intended participant–robot relational framing.
}

{However, this constraint may have undermined ecological validity. 
Some participants disclosed that to some degree, it felt unnatural, and without the constraint they would have interacted differently
(\eg holding, squeezing, or simply observing).
Indeed, we observed some occasionally engaged in these interactions while primarily stroking.
}

{Similarly, some common experiment control measures (repeated neutralizations, restricted motion from sensor tethers) inculcated a lab-study atmosphere despite a deliberate effort to create a living-room aesthetic. 
}

These choices likely dampened rather than inflated the intervention effect. 
Embodied and enactive emotion theories~\cite{di2024enactive} 
suggest that granting more agency over an interaction may promote meaningful engagement; this would potentially enhance regulatory outcomes (with caveats of construct validity).
{Meanwhile, there is room to increase participant comfort and better approximate real-world interactions.
}

\subsection{Future Work}
\label{6:conc:FW}

\lowHeading{Individual Differences in {perceived} ER {Support}}
Participants varied considerably in how they engaged with the \chora’s behaviors, particularly in higher-order processes such as engagement, social bonding, sense-making, and memory recollection. 
While looped behaviors and liveliness were broadly perceived as calming and comforting, for some, they elicited disengagement or discomfort. 
These variations align with broader ER research~\cite{chen2021emotion}
, which emphasizes the role of personal history, dispositional traits, and situational context in shaping regulatory responses.
Building on these insights, we plan to examine how individual differences influence interaction trajectories and susceptibility, seek patterns of such differences,  and propose how \chora\ systems could be tailored for such profiles.

\lowHeading{{Responsive} Behaviors}
We conservatively employed looped (non-interactive) \chora\ behaviors,  
to isolate the effects of haptic sensation from those arising from responsive behaviors.
With this foundational condition confirmed, we can examine whether real-time interactivity tailored to facilitate specific ER strategies could have even greater leverage,
a direction also echoed by several participants.
This opens a broader research space for contextual design, co-created meaning, and agentic evolution. 
While such dynamics have been explored in social robotics through visual and verbal modalities~\cite{feng2022context}, they remain a frontier for affective haptics~\cite{vyas_descriptive_2023}.

Evaluating the emergent outcomes of dyadic agent touch interactions may require a separate evaluation framework and long-term data collection methods focused on trajectories, interaction history, and agent evolution~\cite{butler_emotional_2013}, an exciting direction for future research.

\lowHeading{Evaluating the \chora\ to Support Reappraisal}
{Although this study focused on validating \chora’s generalized  ER facilitation, we aim to further investigate its role in supporting cognitive reappraisal,
by building on cumulative effects that emerge through the successful facilitation of other ER strategies.}
Our theoretical framework~\cite{vyas:2024:happraisalModel} outlines two operational pathways 
through which \chora\ can facilitate cognitive reappraisal.
{With preliminary evidence that \chora\ facilitates all ER strategies, we are now positioned to assess its more direct role in reappraisal and dive deeper.}

\lowHeading{Facilitating Interconnected ER Strategies}
Emotion regulation is rarely a discrete or singular process; users often engage multiple strategies in sequence or combination, with one facilitating the onset of another.
Our findings suggest that successful support may involve multi-sensory coherence (\eg breathing-like movement and soft form), the opportunity for self-orchestrated interaction, and integration of subtle cues that scaffold shifts in attention or sense-making. 
To better align with individual ER styles and in-situ needs, \chora-like systems  may benefit from personalization features such as custom narratives to aid initial social bonding and its evolution, adjustable behaviors based on emotional states and in-the-moment need, or adaptive responses informed by touch and physiology. 
Future work can also explore multisensory levers, such as temperature, ambient sound, or scent.

\section*{Acknowledgments}
Thanks to the Natural Sciences and Engineering Research Council of Canada (NSERC) and Estonian Centre of Excellence for Well-Being Sciences ``EstWell'' (Estonian Ministry of Education and Research grant  TK218 ) for funding this work. Human user research was conducted under UBC Ethics \#H15-02611. 
We thank the contributions from Happraisal Project Team Members: Labella Li, Avi Sharma, Huron Yin, Felicia Yin, Negin Hashemzadeh, and other SPIN Lab members for their support towards this project and manuscript.

\bibliographystyle{IEEEtran}
\bibliography{references}

\vspace{-2.5\baselineskip}  

\begin{IEEEbiographynophoto}{Preeti Vyas}
is a Ph.D. Candidate in Computer Science at the University of British Columbia. 
Her research interest lies in the areas of affective haptics, social robotics, and human-centred design, with a focus on designing technology for facilitating emotion regulation and mental well-being.
She holds a Master's in Electrical Engineering from McGill University and a B.Tech. in Electronics and Communication Engineering from NIT-Bhopal. 
\end{IEEEbiographynophoto}

\vspace{-2.7\baselineskip}  

\begin{IEEEbiographynophoto}{Bereket Guta}
is a Ph.D. student in Computer Science at the University of British Columbia. He holds a Bachelor's degree in Engineering Physics from the same institution. His current research focuses on the intersection of social robotics and affective haptic behaviors. He is particularly interested in applying principles from control theory and cognitive science to develop intuitive and natural robot behaviors that support human needs.
\end{IEEEbiographynophoto}

\vspace{-2.7\baselineskip}  

\begin{IEEEbiographynophoto}{Tim G. Zhou}
is a Ph.D. track Master’s student in Computer Science at the University of British Columbia.
He holds a Bachelor’s degree in Combined Honours in Computer Science and Statistics from the same institution.
His research interests span human-computer interaction and machine learning, with a current focus on uncertainty quantification in deep learning.
\end{IEEEbiographynophoto}

\vspace{-2.7\baselineskip}  

\begin{IEEEbiographynophoto}{Noor Naila Himam}
is a recent graduate of the University of British Columbia, where she completed a B.Sc. in the Combined Major in Science, specializing in Computer Science, Life Science, and Environmental Sciences. Her research interests span human-computer interaction. She currently works as a research assistant, contributing to studies on affective haptics.
\end{IEEEbiographynophoto}

\vspace{-2.7\baselineskip}  

\begin{IEEEbiographynophoto}{Andero Uusberg}
is a Professor of Affective Psychology and PI of the Estonian Center of Excellence for Well-Being Sciences at the University of Tartu. With a PhD in psychology, Andero investigates how emotions and other affective states arise from the predictive computations of the mind and how these states can be adjusted using reappraisal and other emotion regulation strategies.   
\end{IEEEbiographynophoto}

\vspace{-2.7\baselineskip}  

\begin{IEEEbiographynophoto}{Karon E. MacLean}
is Professor of Computer Science and a Canada Research Chair in Interactive Human Systems Design at the University of British Columbia in Canada. 
With degrees in biology and mechanical engineering (Stanford, MIT), her research involves haptic technology and affective haptics in human–computer interaction and human–robot interaction.
\end{IEEEbiographynophoto}

\vfill
\newpage


\newpage













\appendices

\section{Custom Questionnaires}
\label{app:custom-qs}
\renewcommand{\thesection}{\Alph{section}}

\subsection*{ \textbf{A.1 Affective Object Orientation Questionnaire (Pre-Test)}}
\label{app:aooq}
\addcontentsline{toc}{section}{A.1: Affective Object Orientation Questionnaire (Pre-Test)}

\noindent How well do the following statements describe you? In the context of this question, a personal object is something you own.
\vspace{2mm}

\begin{quote}
I have a tendency to {fidget}. \\
I have a tendency to develop {emotional attachment} to personal objects.\\
I have a tendency to {value design and aesthetics when choosing} personal objects.\\
I am likely to own a {personal object that doesn’t serve any functional needs}.
\end{quote}

\vspace{2mm}
\noindent \textit{Response scale: 1 = Strongly disagree, 2 = Somewhat disagree, 3 = Neither agree nor disagree, 4 = Somewhat agree, 5 = Strongly agree}
\renewcommand{\thesection}{\Alph{section}}

\subsection*{ \textbf{A.2 Internal Dialogue Reflection (Post-Test)}}
\label{app:idr}
\addcontentsline{toc}{section}{A.2: Internal Dialogue Reflection (Post-Test)}
\noindent 
Think back over the last hour and answer the following questions.
\vspace{2mm}

\noindent{Q1. During the experiment, I thought about:} \\
… this robot and other aspects of this experiment \\
… people and relationships \\
… my own thoughts, feelings, and bodily sensations \\
… the future \\
… the past \\
… the here and now \\
… nothing \\
… something else (please specify): 
\vspace{2mm}

\noindent\textit{Response Scale: 1 = Not at all, 2 = Rarely, 3 = Sometimes, 4 = Often, 5 = All the time} \\

\vspace{2mm}

\noindent{Q2. To what degree were these thoughts demanding your attention?} 

\vspace{2mm}
\noindent\textit{Response Scale: 1 = Not at all, 2 = Slightly, 3 = Moderately, 4 = Very Much, 5 = Completely} \\

\vspace{2mm}

\noindent{Q3. Please rate the polarity of your thoughts.} \vspace{2mm}
\begin{quote}
Right now, I’d describe my most recent thoughts as: \\
During the study, I’d describe my thoughts as: \\
At the start of the study, I’d describe my thoughts as: 
\end{quote}

\vspace{2mm}
\noindent\textit{Response Scale: 1 = Very Negative, 2 = Negative, 3 = Neutral, 4 = Positive, 5 = Very Positive} \\

\vspace{2mm}

\noindent{Q4. To what degree do you think the CHORA (in any of its conditions) had an impact on the trajectory of your thoughts?} 

\vspace{2mm}
\begin{quote}
The CHORA had no impact. \\
The CHORA had a positive impact. \\
The CHORA had a negative impact.  \\
\end{quote}

\noindent\textit{Response Scale: 1 = Strongly Disagree, 2 = Disagree, 3 = Neutral, 4 = Agree, 5 = Strongly Agree} \\

\vspace{2mm}

\noindent Q5. Can you describe this impact? You can list adjectives or elaborate on your experience. [open-ended] \\

\subsection*{ \textbf{A.3 CHORA ER Facilitation Questionnaire (Post-Test)}}
\label{app:ERfac}
\addcontentsline{toc}{section}{A.3: CHORA Emotion Regulation Facilitation Questionnaire (Post-Test)}
\noindent Please reflect on your interaction with the CHORA throughout the session and respond to the following questions based on your thoughts, feelings, and sensory experiences. \\

\vspace{1mm}
\noindent{1. Interacting with the CHORA directed my attention away from:} 

\noindent
... my thoughts and feelings \\
... my bodily sensations

\vspace{2mm}

\noindent{2. Interacting with the CHORA felt:}

\noindent
… pleasant \\
… familiar

\vspace{2mm}

\noindent{3. Interacting with the CHORA made me feel:}

\noindent
… comforted \\
… calm \\
… secure and safe \\
… happy and excited \\
… uneasy \\
… unsettled \\
… anxious \\
… unsafe \\
… gloomy

\vspace{2mm}

\noindent{4. Interacting with the CHORA caused me to think of:}

\noindent
… a pet \\
… a loved one \\
… a comfort object \\
… myself

\vspace{2mm}

\noindent{5. Interacting with the CHORA made me feel a sense of:}

\noindent
… a sense of connection \\
… a sense of reliance \\
… a sense of social support \\
… a sense of being grounded or centred

\vspace{2mm}

\noindent{6. Interacting with the CHORA helped me to:}

\noindent
… appreciate things as they are \\
… see things in a positive light \\
… see things in perspective\\

\noindent\textit{Response scale: 1 = Strongly disagree, 2 = Somewhat disagree, 3 = Neither agree nor disagree, 4 = Somewhat agree, 5 = Strongly agree}

\newpage
\onecolumn
\section{Haptic Behavior}
\label{app: haptic behavior}


\begin{figure*}[htb]
\centering
\includegraphics[width=0.75\columnwidth]{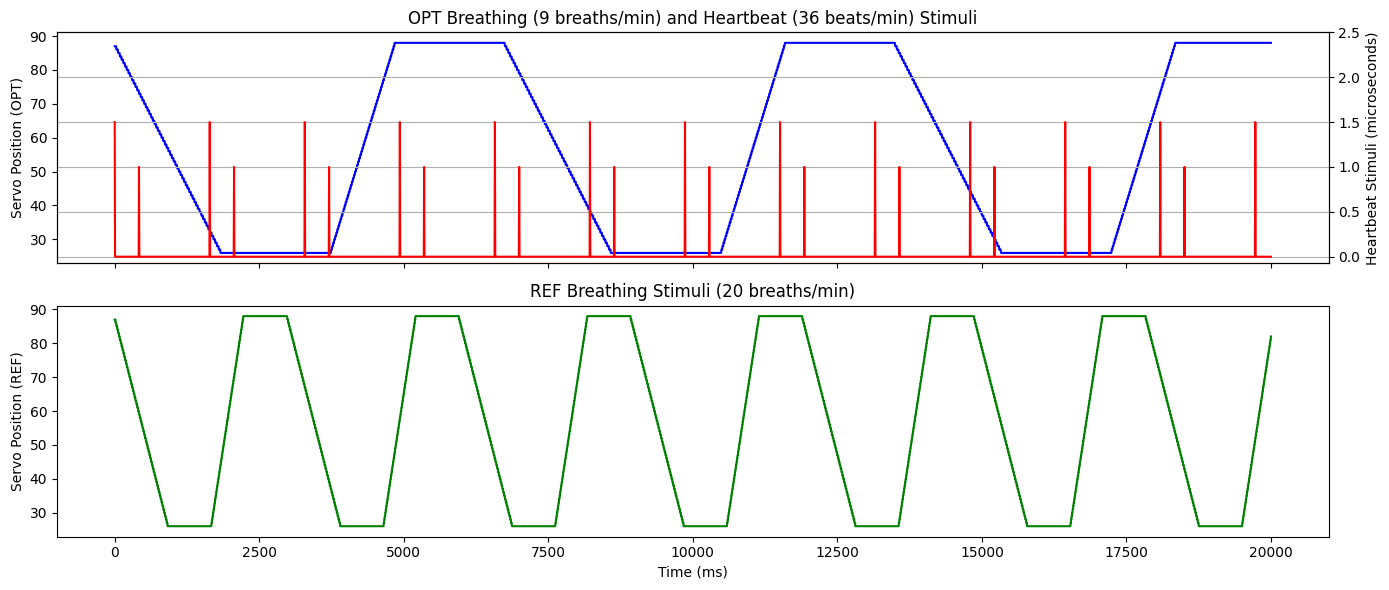}
\caption{Haptic behavior of the \chora\ for \OPT\ and \REF\ conditions (20s snippet). OPT condition has both heartbeat and breathing stimuli. Heartbeat stimuli are represented by an impulse signal in red with its intensity (duration) on the y-axis. The breathing for both REF (green) and OPT (blue) followed a trapezoidal wave with different parametrizations.}
\label{fig:app:stimuli}
\end{figure*}
\newpage

\section{Experiment Condition Order}
\label{app: condition order}


\begin{table*}[htb]
\centering
\caption{Randomized experimental condition order; p means practice round, c1, c2, c3 represent conditions in presentation order, b means block, there were three blocks b1, b2, b3. We highlight one set of condition order to showcase its occurrence.}
\label{tab:appendix:condition-orders}
\resizebox{\textwidth}{!}{%
\begin{tabular}{c cccccccccccccccc}
\toprule
pno & pbaseline & pc1 & pc2 & pc3 & b1baseline & b1c1 & b1c2 & b1c3 & b2baseline & b2c1 & b2c2 & b2c3 & b3baseline & b3c1 & b3c2 & b3c3 \\
\midrule
P1  & ABS & INA & OPT & REF & ABS & \cellcolor{yellow} OPT & \cellcolor{yellow} REF & \cellcolor{yellow} INA & ABS & REF & INA & OPT & ABS & REF & OPT & INA \\
P2  & ABS & REF & OPT & INA & ABS & OPT & INA & REF & ABS & INA & REF & OPT & ABS & REF & INA & OPT \\
P3  & ABS & REF & INA & OPT & ABS & INA & OPT & REF & ABS & \cellcolor{yellow} OPT & \cellcolor{yellow} REF & \cellcolor{yellow} INA & ABS & INA & REF & OPT \\
P4  & ABS & OPT & INA & REF & ABS & INA & REF & OPT & ABS & REF & OPT & INA & ABS & INA & OPT & REF \\
P5  & ABS & \cellcolor{yellow} OPT & \cellcolor{yellow} REF & \cellcolor{yellow} INA & ABS & REF & INA & OPT & ABS & INA & OPT & REF & ABS & OPT & INA & REF \\
P6  & ABS & INA & REF & OPT & ABS & REF & OPT & INA & ABS & OPT & INA & REF & ABS & \cellcolor{yellow} OPT & \cellcolor{yellow} REF & \cellcolor{yellow} INA \\
P7  & ABS & INA & OPT & REF & ABS & \cellcolor{yellow} OPT & \cellcolor{yellow} REF & \cellcolor{yellow} INA & ABS & REF & INA & OPT & ABS & REF & OPT & INA \\
P8  & ABS & REF & OPT & INA & ABS & OPT & INA & REF & ABS & INA & REF & OPT & ABS & REF & INA & OPT \\
P9  & ABS & REF & INA & OPT & ABS & INA & OPT & REF & ABS & \cellcolor{yellow} OPT & \cellcolor{yellow} REF & \cellcolor{yellow} INA & ABS & INA & REF & OPT \\
P10 & ABS & OPT & INA & REF & ABS & INA & REF & OPT & ABS & REF & OPT & INA & ABS & INA & OPT & REF \\
P11 & ABS & \cellcolor{yellow} OPT & \cellcolor{yellow} REF & \cellcolor{yellow} INA & ABS & REF & INA & OPT & ABS & INA & OPT & REF & ABS & OPT & INA & REF \\
P12 & ABS & INA & REF & OPT & ABS & REF & OPT & INA & ABS & OPT & INA & REF & ABS & \cellcolor{yellow} OPT & \cellcolor{yellow} REF & \cellcolor{yellow} INA \\
P13 & ABS & INA & OPT & REF & ABS & \cellcolor{yellow} OPT & \cellcolor{yellow} REF & \cellcolor{yellow} INA & ABS & REF & INA & OPT & ABS & REF & OPT & INA \\
P14 & ABS & REF & OPT & INA & ABS & OPT & INA & REF & ABS & INA & REF & OPT & ABS & REF & INA & OPT \\
P15 & ABS & REF & INA & OPT & ABS & INA & OPT & REF & ABS & \cellcolor{yellow} OPT & \cellcolor{yellow} REF & \cellcolor{yellow} INA & ABS & INA & REF & OPT \\
P16 & ABS & OPT & INA & REF & ABS & INA & REF & OPT & ABS & REF & OPT & INA & ABS & INA & OPT & REF \\
P17 & ABS & \cellcolor{yellow} OPT & \cellcolor{yellow} REF & \cellcolor{yellow} INA & ABS & REF & INA & OPT & ABS & INA & OPT & REF & ABS & OPT & INA & REF \\
P18 & ABS & INA & REF & OPT & ABS & REF & OPT & INA & ABS & OPT & INA & REF & ABS & \cellcolor{yellow} OPT & \cellcolor{yellow} REF & \cellcolor{yellow} INA \\
P19 & ABS & INA & OPT & REF & ABS & \cellcolor{yellow} OPT & \cellcolor{yellow} REF & \cellcolor{yellow} INA & ABS & REF & INA & OPT & ABS & REF & OPT & INA \\
P20 & ABS & REF & OPT & INA & ABS & OPT & INA & REF & ABS & INA & REF & OPT & ABS & REF & INA & OPT \\
P21 & ABS & REF & INA & OPT & ABS & INA & OPT & REF & ABS & \cellcolor{yellow} OPT & \cellcolor{yellow} REF & \cellcolor{yellow} INA & ABS & INA & REF & OPT \\
P22 & ABS & OPT & INA & REF & ABS & INA & REF & OPT & ABS & REF & OPT & INA & ABS & INA & OPT & REF \\
P23 & ABS & \cellcolor{yellow} OPT & \cellcolor{yellow} REF & \cellcolor{yellow} INA & ABS & REF & INA & OPT & ABS & INA & OPT & REF & ABS & OPT & INA & REF \\
P24 & ABS & INA & REF & OPT & ABS & REF & OPT & INA & ABS & OPT & INA & REF & ABS & \cellcolor{yellow} OPT & \cellcolor{yellow} REF & \cellcolor{yellow} INA \\
P25 & ABS & INA & OPT & REF & ABS & \cellcolor{yellow} OPT & \cellcolor{yellow} REF & \cellcolor{yellow} INA & ABS & REF & INA & OPT & ABS & REF & OPT & INA \\
P26 & ABS & REF & OPT & INA & ABS & OPT & INA & REF & ABS & INA & REF & OPT & ABS & REF & INA & OPT \\
P27 & ABS & REF & INA & OPT & ABS & INA & OPT & REF & ABS & \cellcolor{yellow} OPT & \cellcolor{yellow} REF & \cellcolor{yellow} INA & ABS & INA & REF & OPT \\
P28 & ABS & OPT & INA & REF & ABS & INA & REF & OPT & ABS & REF & OPT & INA & ABS & INA & OPT & REF \\
P29 & ABS & \cellcolor{yellow} OPT & \cellcolor{yellow} REF & \cellcolor{yellow} INA & ABS & REF & INA & OPT & ABS & INA & OPT & REF & ABS & OPT & INA & REF \\
P30 & ABS & INA & REF & OPT & ABS & REF & OPT & INA & ABS & OPT & INA & REF & ABS & \cellcolor{yellow} OPT & \cellcolor{yellow} REF & \cellcolor{yellow} INA \\
\bottomrule
\end{tabular}%
}
\end{table*}
\newpage

\section{Demographics}
\label{app: demographics}

\begin{table*}[htb]
\caption{Demographic data of the study participants. Enclosed in parentheses is the count of participants who responded to the respective field. * denotes fields with multiple responses.}
\label{tab:app:demographicsGeneral}
\centering
\small
\begin{tabular}{llll}
\toprule
\textbf{Demographic}                     & \textbf{Response Categories}             & \textbf{Count} & \textbf{Percentage (\%)}        \\ 
\midrule

\textbf{Gender (n=29)}                          
& Female                       & 15  & 50.00 \\
& Male                         & 14  & 46.67 \\

\\[-0.7em]
\textbf{Age (n=30)}                            
& 18--24                        & 12  & 40.00 \\
& 25--35                        & 17  & 56.67 \\
& 35--45                        & 1   & 3.33  \\

\\[-0.7em]
\textbf{Education (n=30)}                      
& High School                   & 2   & 6.67  \\
& Some college credit           & 1   & 3.33  \\
& Bachelor's                    & 17  & 56.67 \\
& Master's                      & 9   & 30.00 \\
& Doctorate                     & 1   & 3.33  \\

\\[-0.7em]
\textbf{Lived Experiences* (n=28)}             
& Student                       & 20  & 66.67 \\
& Person of Colour              & 13  & 43.33 \\
& Immigrant                     & 11  & 36.67 \\
& LGBTQIA+                      & 4   & 13.33 \\
& Low-Income                    & 4   & 13.33 \\
& Neurodivergent                & 4   & 13.33 \\
& Medical Condition             & 2   & 6.67  \\
& Minority Religion             & 3   & 10.00 \\
& Violence Survivor             & 1   & 3.33  \\
& Marginalised Gender Identity  & 1   & 3.33  \\
& Domestic Violence Survivor    & 1   & 3.33  \\
& Homelessness                  & 2   & 6.67  \\
& Not applicable                & 1   & 3.33  \\

\\[-0.7em]
\textbf{Cultural Background* (n=27)}           
& East Asian                    & 12  & 40.00 \\
& European/White                & 7   & 23.33 \\
& South Asian                   & 5   & 16.67 \\
& Black/African                 & 1   & 3.33  \\
& Middle Eastern                & 1   & 3.33  \\
& Self-described                & 1   & 3.33  \\

\\[-0.7em]
\textbf{Has a Pet (n=30)}              
& Yes                           & 18  & 60.00 \\
& No                            & 12  & 40.00 \\

\\[-0.7em]
\textbf{Handedness (n=30)}
& Right                         & 28  & 93.33 \\
& Left                          & 1   & 3.33  \\
& Ambidextrous                  & 1   & 3.33  \\

\bottomrule
\end{tabular}

\vspace{1em}

\begin{tabular}{lllllll}
\toprule
\textbf{Scale} & \textbf{Subscale} & \textbf{Mean} & \textbf{Median} & \textbf{Std. Dev.} & \textbf{Min} & \textbf{Max} \\
\midrule

\multirow{5}{*}{\textbf{BIG-5 TRAITS (n=30)}}
& Extraversion      & 6.00  & 6.00  & 1.95 & 2 & 9 \\
& Agreeableness     & 7.30  & 8.00  & 1.64 & 4 & 10 \\
& Conscientiousness & 6.77  & 7.00  & 1.76 & 3 & 10 \\
& Neuroticism       & 6.73  & 6.50  & 2.08 & 4 & 10 \\
& Openness          & 7.10  & 7.00  & 1.42 & 5 & 10 \\

\\[-0.7em]
\multirow{2}{*}{\textbf{ERQ (n=30)} }
& Reappraisal Score & 3.45  & 3.67  & 0.72 & 1.00 & 5.00 \\
& Suppression Score & 2.83  & 2.88  & 0.97 & 1.25 & 4.75 \\

\\[-0.7em]
\multirow{3}{*}{\textbf{DASS-21 (n=30)}}
& Depression        & 9.60  & 8.00  & 8.01 & 0 & 28 \\
& Anxiety           & 10.07 & 9.00  & 6.27 & 0 & 28 \\
& Stress            & 14.13 & 12.00 & 9.26 & 0 & 34 \\

\bottomrule
\end{tabular}
\end{table*} 
\newpage

\section{Statistics}
\label{app: statistics}

\begin{table*}[htb]
\centering
\caption{Means, Standard Deviations, and Offset Values for Collected Metrics}
\label{tab:s5:quant-descriptive}
\begin{tabular}{lcccccccccccccc}
\toprule
\multicolumn{1}{l}{} & \multicolumn{8}{c}{\textbf{Absolute Values}} & \multicolumn{6}{c}{\textbf{Offset Values} (\textsc{abs}-cond)} \\
\cmidrule(lr){2-9} \cmidrule(lr){10-15}
 \textbf{Conditions} & \multicolumn{2}{c}{ABS} & \multicolumn{2}{c}{INA} & \multicolumn{2}{c}{REF} & \multicolumn{2}{c}{OPT} & \multicolumn{2}{c}{$\Delta_\textsc{{INA}}$} & \multicolumn{2}{c}{$\Delta_\textsc{REF}$} & \multicolumn{2}{c}{$\Delta_\textsc{OPT}$} \\
\cmidrule(lr){2-3} \cmidrule(lr){4-5} \cmidrule(lr){6-7} \cmidrule(lr){8-9} \cmidrule(lr){10-11} \cmidrule(lr){12-13} \cmidrule(lr){14-15}
& mean & std & mean & std & mean & std & mean & std & mean & std & mean & std & mean & std \\
\midrule

SAM-V (29)  & 5.68 & 1.26 & 5.69 & 1.58 & 6.55 & 1.53 & 6.55 & 1.33 & 
0.00 & 1.51 & -0.86 & 1.36 & -0.86 & 1.25 \\ 

SAM-A (29)  & 2.61 & 1.28 & 2.45 & 1.15 & 2.55 & 1.45 & 2.34 & 1.40 & 
0.52 & 1.35 & 0.41 & 1.64 & 0.62 & 1.45 \\ 

SAM-D (29)  & 5.06 & 2.04 & 5.52 & 2.10 & 5.45 & 2.29 & 5.24 & 2.18 & 
-0.55 & 1.4 & -0.48 & 1.39 & -0.28 & 1.39 \\ 
 
STAI-6 (30) & 9.72 & 2.57 & 9.23 & 1.85 & 8.67 & 2.01 & 8.5 & 1.83 & 
1.13 & 2.57 & 1.87 & 2.22 & 1.7 & 2.12 \\ 

HR (29)     & 74.10 & 10.57 & 75.66 & 10.39 & 74.67 & 10.46 & 74.62 & 
10.23 & -1.55 & 2.60 & -0.56 & 2.69 & -0.52 & 3.13 \\

GSR (30)  & 18.43 & 11.95 & 22.87 & 14.59 & 21.86 & 13.93 & 22.37 & 
14.26 & -4.44 & 6.22 & -3.43 & 5.31 & -3.94 & 5.57 \\

\bottomrule
\end{tabular}
\end{table*}
\begin{table*}[htb]
\centering
\caption{Shapiro-Wilk Normality Test Results for SAM, HR, GSR, and STAI-6 Measures (Delta Conditions) p$>$0.05 than normal distributions}
\label{tab:app:normality-checkDelta}
\begin{tabular}{lllcc}
\toprule
\textbf{Measure} & \textbf{Condition} & \textbf{p-value} & \textbf{Normality} \\
\midrule
\multirow{3}{*}{SAM-V} 
  & $\Delta_{\textsc{INA}}$ & 0.01 & No \\
  & $\Delta_{\textsc{OPT}}$ & 0.02 & No \\
  & $\Delta_{\textsc{REF}}$ & 0.02 & No \\
\cmidrule{2-4}
  & \multicolumn{2}{r}{\textit{Statistical Test:}} & \multicolumn{1}{l}{\textit{Friedman (post-hoc Wilcoxon)}} \\
\midrule
\multirow{3}{*}{SAM-A} 
  & $\Delta_{\textsc{INA}}$ & 0.08 & Yes \\
  & $\Delta_{\textsc{OPT}}$ & 0.03 & No \\
  & $\Delta_{\textsc{REF}}$ & 0.21 & Yes \\
\cmidrule{2-4}
  & \multicolumn{2}{r}{\textit{Statistical Test:}} & \multicolumn{1}{l}{\textit{Friedman (post-hoc Wilcoxon)}} \\
\midrule
\multirow{3}{*}{SAM-D} 
  & $\Delta_{\textsc{INA}}$ & 0.01 & No \\
  & $\Delta_{\textsc{OPT}}$ & 0.03 & No \\
  & $\Delta_{\textsc{REF}}$ & 0.07 & Yes \\
\cmidrule{2-4}
  & \multicolumn{2}{r}{\textit{Statistical Test:}} & \multicolumn{1}{l}{\textit{Friedman (post-hoc Wilcoxon)}} \\
\midrule
\multirow{3}{*}{STAI-6} 
  & $\Delta_{\textsc{INA}}$ & 0.33 & Yes \\
  & $\Delta_{\textsc{OPT}}$ & 0.16 & Yes\\
  & $\Delta_{\textsc{REF}}$ & 0.32 & Yes \\
\cmidrule{2-4}
  & \multicolumn{2}{r}{\textit{Statistical Test:}} & \multicolumn{1}{l}{\textit{ANOVA (post-hoc t-tests)}} \\
\midrule
\multirow{6}{*}{HR} 
  & $\Delta_{\textsc{INA}}$ & 0.34 & Yes \\
  & $\Delta_{\textsc{OPT}}$ & 0.37 & Yes \\
  & $\Delta_{\textsc{REF}}$ & 0.17 & Yes \\
\cmidrule{2-4}

  & \multicolumn{2}{r}{\textit{Statistical Test:}} & \multicolumn{1}{l}{\textit{ANOVA (post-hoc t-tests)}} \\
\midrule
\multirow{6}{*}{GSR} 
  & $\Delta_{\textsc{INA}}$ & 0.00 & No \\
  & $\Delta_{\textsc{OPT}}$ & 0.00 & No \\
  & $\Delta_{\textsc{REF}}$ & 0.00 & No \\
\cmidrule{2-4}
& \multicolumn{2}{r}{\textit{Statistical Test:}} & \multicolumn{1}{l}{\textit{Friedman (post-hoc Wilcoxon)}} \\
\bottomrule
\end{tabular}
\end{table*}
\newpage

\section{\rev{Physiological Data for All Rounds}}
\label{app: carry-over}


\begin{figure*}[h]
\centering
\includegraphics[width=\columnwidth]{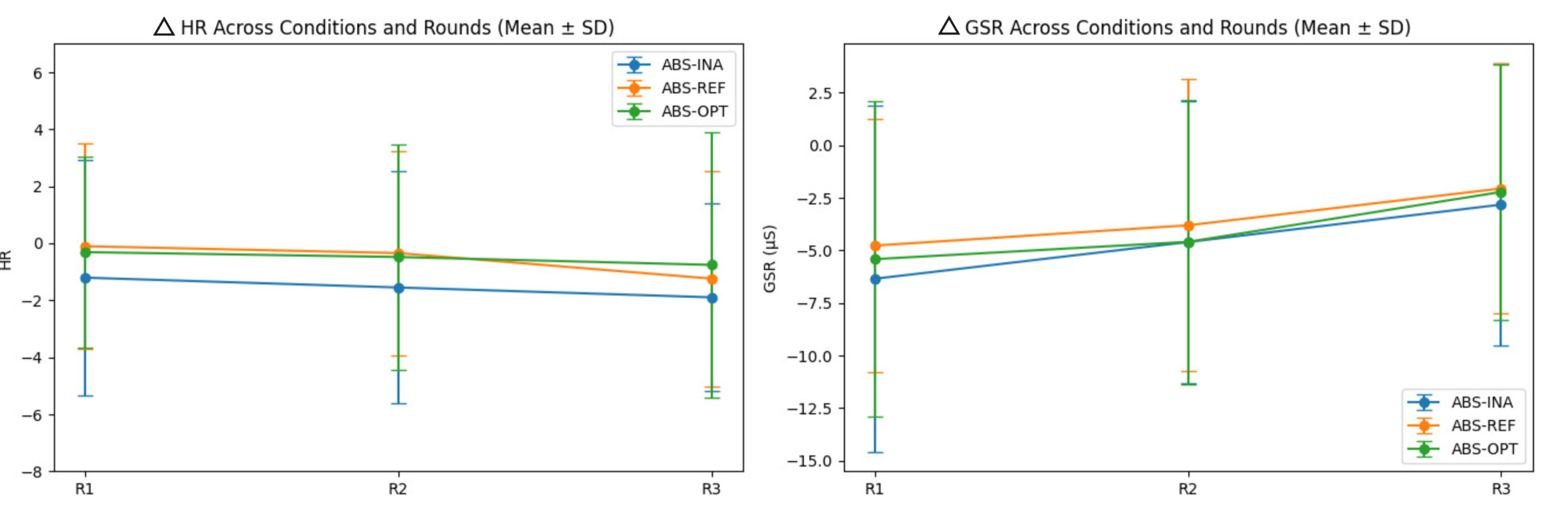}
\caption{\rev{Physiological responses to the \chora\ interaction by rounds. (A) Mean changes in Heart Rate (\HR) and (B) Mean changes in Galvanic Skin Response (\GSR) across three conditions (\dINA, \dREF, \dOPT) and three experimental rounds (R1–R3). 
Error bars represent standard deviation.}}
\label{fig:app:cond by rounds}
\end{figure*}

\begin{figure*}[htb]
\centering
\includegraphics[width=\columnwidth]{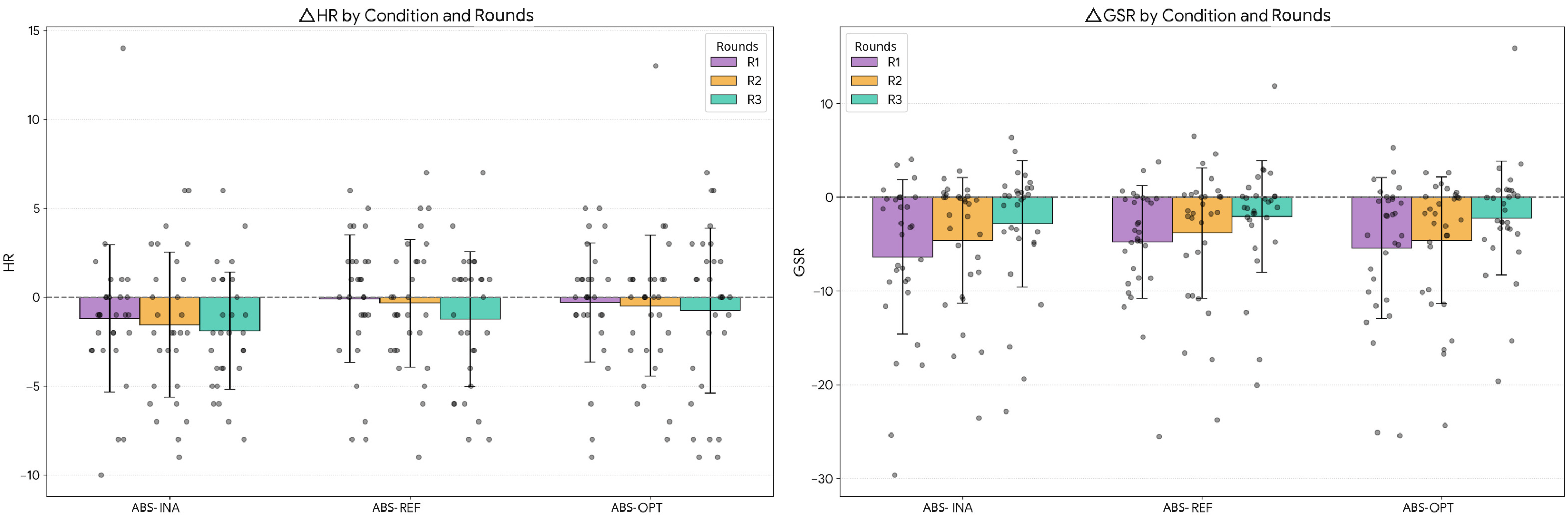}
\caption{\rev{Physiological responses to the \chora\ interaction by conditions. (A) Mean changes in Heart Rate (\HR) and (B) Mean changes in Galvanic Skin Response (\GSR) across three conditions (\dINA, \dREF, \dOPT) and three experimental rounds (R1–R3). 
In both plots, bars indicate the mean change from baseline, error bars represent standard deviation, and individual data points are overlaid to illustrate the range of participant responses. 
Higher values (less negative or above zero) in both metrics generally reflect a physiological state of relaxation or reduced arousal.}}
\label{fig:app:rounds by cond}
\end{figure*}
\newpage

\section{\rev{Our Robot Prototype \vs Haptic Creature}}
\label{app: robot}
\begin{table}[h!]
\centering
\begin{tabular}{lcc}
\toprule
\rev{\textbf{Feature}} & \rev{\textbf{The CHORA Prototype}} & \rev{\textbf{Haptic Creature}} \\
\toprule
 & \includegraphics[width=0.19\textwidth]{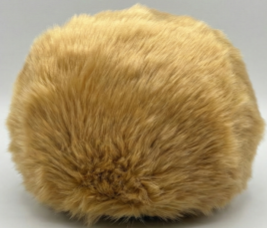} & \includegraphics[width=0.25\textwidth]{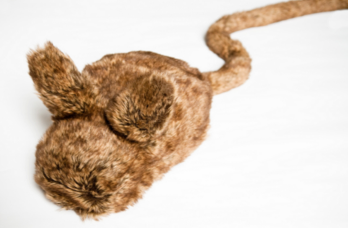} \\
\toprule
\multicolumn{3}{c}{\rev{\textbf{Form Factor and Body}}} \\
\midrule
\rev{Shape} & \rev{Spherical / Ball-like} & \rev{Lap-sized Animal}  \\
\rev{Appearance} & \rev{Abstract zoomorphic} & \rev{Literal zoomorphic with ears, face, and a tail} \\
\rev{Size} & \rev{Compact (Max dimension: 24 cm)} & \rev{Larger \& Elongated (Max dimension: 33 cm)} \\
\rev{Weight} & \rev{Lightweight (800 g)} & \rev{Heavy (2.5 kg)} \\
\rev{Outer Shell} & \rev{Faux fur over 1.5 cm foam} & \rev{Faux fur over 0.25" Silicon Rubber (``Dragon Skin'')} \\
\rev{Inner Structure} & \rev{Ribbit wooden ribcage covered in thick foam} & \rev{A custom fiberglass laminate ``turtle'' shell} \\

 & \includegraphics[width=0.19\textwidth]{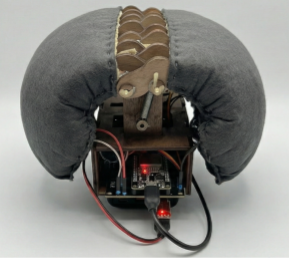} & \includegraphics[width=0.25\textwidth]{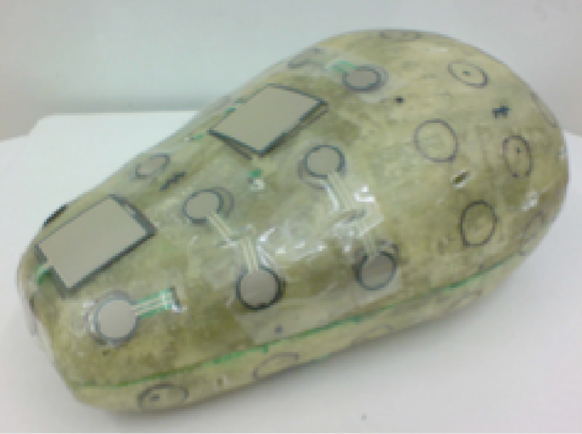} \\

\midrule
\multicolumn{3}{c}{\rev{\textbf{Electronics and Sensing}}} \\
\midrule
\rev{Connectivity} & \rev{Untethered (Wi-Fi, Battery)} & \rev{Tethered (USB + External Power)} \\
\rev{Microcontroller} & \rev{ESP32} & \rev{Microchip PIC18F87J50 USB Board} \\
\rev{Sensing} & \rev{IMU, Hall Effect, Custom Touch Sensors} & \rev{60 Force Sensing Resistors (FSRs)} \\

\midrule
\multicolumn{3}{c}{\rev{\textbf{Interaction Capabilities}}} \\
\midrule
\rev{Interactions} & \rev{Breathing + Heartbeat} & \rev{Breathing + Purring + Ear Stiffening} \\
\rev{Tactile} & \rev{Breathing: Hitec HS-485HB Servo} & \rev{Breathing: Hitec HSR-5980SG Servo (T-lever)} \\
\rev{Vibrotactile} & \rev{Heartbeat: Titan Haptics DRAKE LF} & \rev{Purring: Maxon RE025 DC Motor} \\
\rev{Others} & \rev{None} & \rev{Ears: Hitec HS322HD Servos (pneumatic)} \\

\midrule
\multicolumn{3}{c}{\rev{\textbf{Behaviors Used in the Study}}} \\
\midrule
& \rev{Ours: Both breathing and heartbeat activated} & \rev{Sefidgar~\etal: Only breathing activated} \\
\bottomrule
\end{tabular}
\caption{\rev{Comparison of features between the \chora\ Prototype and a Haptic Creature from previous work.}}
\label{tab:chora-comparison}
\end{table}

\end{document}